\documentclass[12pt,preprint]{aastex} 
\tighten 

\begin{document}

%\received{}
\accepted{May 22, 2009, Astrophysical Journal}

\title{A FUSE Survey of the Rotation Rates of Very Massive Stars in the Small and Large Magellanic Clouds}

\author{Laura R. Penny} 
\affil{Department of Physics and Astronomy \\
       The College of Charleston \\
       Charleston, SC 29424 \\
       pennyl@cofc.edu}
\author{Douglas R. Gies}   
\affil{ Center for High Angular Resolution Astronomy and \\ 
       Department of Physics and Astronomy,\\
       Georgia State University, 
       P.O. Box 4106,
       Atlanta, GA 30302-4106 \\ 
       gies@chara.gsu.edu}

%\comment{}

%%%%%%%%%%%%%%%%%%%%%%%%%%%%%%%%%%%%%%%%%%%%%%%%%%%%%%%%%%%

\begin{abstract} 
We present projected rotational velocity values for  97 Galactic, 55 SMC, and 106 LMC O-B type stars from archival FUSE observations.  The evolved and unevolved samples from each environment are compared through the Kolmogorov-Smirnov test to determine if the distribution of equatorial rotational velocities is metallicity dependent for these massive objects.  Stellar interior models predict that massive stars with SMC metallicity will have significantly reduced angular momentum loss on the main sequence compared to their Galactic counterparts.  Our results find some support for this prediction but also show that even at Galactic metallicity, evolved and unevolved massive stars have fairly similar fractions of stars with large $V \sin i$ values.  Macroturbulent broadening that is present in the spectral features of Galactic evolved massive stars is lower in the LMC and SMC samples.  This suggests the processes that lead to macroturbulence are dependent upon metallicity.

\end{abstract}


\pagebreak

\section{Introduction} 

Stellar interior models are the backbone of modern stellar astrophysics.  Their predictions are used in a myriad of applications including determinations of stellar masses from observed luminosities and temperatures, studies of star clusters, population synthesis, stellar nucleosynthesis, chemical evolution, etc.  Many factors are included in these extremely complex interior codes: overshooting, mass loss rates, opacities, convection.  The latest addition to these models has been the inclusion of rotation and its subsequent effects on interior mixing \citep{Mey00, Heg00, Eks08}.  With each additional advance in these codes, observers have been tasked with testing the new models by finding observable diagnostics.  What do the new models predict that can be observed to either agree or disagree with the model?  For massive stars the newest models are very interesting.  Rotation induces interior mixing that changes the luminosities, lifetimes, effective temperatures, and surface abundances of these massive stars.  The combined effect of mass loss and angular momentum in the models results in angular momentum loss; as the mass comes off it takes with it any angular momentum it may have had.  This results in an evolution of surface rotation speed that varies depending upon the amount of mass loss.

The most striking prediction of the new models is the effect of rotation on the internal structure and subsequent luminosity and main sequence lifetime of the star.  The models with rotation predict evolutionary tracks that become more luminous.  This results from two effects: rotational mixing of H-fuel into the convective core and transport of He and other H-burning by-products into the radiative envelope.  The first effect increases the mass of the convective core, while the second lowers the opacity of the envelope.   \citet{Heg00} state, ``As a result of these dependencies of the stellar evolutionary tracks in the HR diagram on the initial rotation rate, a given point in the HR diagram is not uniquely related to a single initial mass, even for core hydrogen burning stars.'' This introduces a serious scatter in the mass-luminosity relation.  In order to determine the evolutionary mass for an object, the luminosity, effective temperature, and the initial equatorial velocity of the star are needed, but this last parameter is indeterminable.  One can attempt to test this prediction of rotationally induced overluminosity.   To test if a single star is overluminous for its mass, first requires knowledge of the mass of the star.  The only direct means to determine the mass is from binary systems.  There are cases where members of a binary system with large equatorial velocities do appear overluminous for their masses \citep{Pen99}.  However other effects can also cause binary stars to appear overluminous, mass transfer being the most common.  Also members of a binary pair exchange spin and orbital angular momentum as they evolve due to tidal interactions, making them poor tests of single star models.

The models also predict how the surface rotation rates should change as the stars evolve.  For solar metallicity stars ($Z_{MW} = 0.020$) the most massive should slow their rotation drastically while still on the main sequence (MS).  There are two competing factors that govern the surface rotation rates: (1)  transfer of angular momentum from the inner parts to the outer ones through meridional circulation and (2) angular momentum loss at the surface of the star through mass loss.  The first of these is present in all rotating stars and is primarily dependent upon the rotation rate: the faster the spin, the more mixing occurs.   The mass loss is from large stellar winds that are driven by line opacities associated with metallic elements.  For stars at lower metallicities (i.e., Small Magellanic Cloud, $Z_{SMC} = 0.004$ and Large Magellanic Cloud, $Z_{LMC} = 0.007$), there are significantly lower numbers of metal atoms and therefore lower line opacity.  Mass loss rates are lower and subsequent angular momentum loss rates are lower too.  The rotation rates of massive stars at lower $Z$ are predicted to remain almost constant during the MS.  For example Figure 10 of \citet{Mey00} shows the evolution of surface rotational velocities as a function of time for  a 20 $M_{\odot}$ star at solar and SMC metallicity with an initial rotation speed of 300 km s$^{-1}$.  By the late main sequence the rotation rate of the Galactic star has decreased to ~140 km s$^{-1}$, while its SMC counterpart still has a rotation rate of ~220 km s$^{-1}$. 

It is interesting to consider what spectral class stars these models refer to, as stars do not come labeled with their initial masses.  The new models predict that stars with initial masses smaller than 15 $M_{\odot}$ do not slow significantly while on the MS even in the Milky Way.  There is no metallicity dependence on their rotational evolution, i.e., a 12 $M_{\odot}$ star at both solar and SMC metallicity have similar behavior.  Our prime targets for testing purposes should be stars that started with 15 or more solar masses.  In addition we see that after the end of the MS, the surface rotational velocities of even these massive stars drop precipitously in all metallicity environments.  A star with an initial mass larger than 15 $M_{\odot}$ spends its MS life as an O-type star (see Fig.~1).  In fact, we see that despite expectations that luminosity classes of giant or supergiant describe post-MS stars, this is not possible for O spectral types.  All O-type stars, regardless of their luminosity class, are core-hydrogen burning objects.  They will eventually become B and A supergiants, but when they do they will have greatly increased radii, and decreased surface rotational velocities.  They also do not exclusively populate these types.  In Figure 1  we over plot the effective temperature line of 14900K, which corresponds to the luminosity class B3 I  \citep{Boh81}.  Even a star with initial mass of 7 $M_{\odot}$ becomes a B3 I \citep[according to][$T_{\rm eff} = 15100$K, $\log g = 3.25$ occurs at age =$ 4.32\times 10^{7}$ yrs for the 7 $M_{\odot}$ track]{Sch92}.  Early B-type supergiants are not an exclusive group and for stars with $M_{ini} \ge 15 M_{\odot}$, they represent stars past the end of the MS.   As such they are not optimal targets for testing the predictions from the new stellar interior models.

There exists observational evidence that the predictions for solar metallicity stars are correct.  Galactic O-type stars do slow down as they evolve along the main sequence \citep{Con77,Pen96,How97}.  There are few broad-lined stars among the more massive and more evolved objects.  The fastest rotators are found among the lower mass O stars.  This agrees well with predictions: that the rotation rates should decrease while still on the MS; that the larger the initial mass, the faster the decrease in rotation; and that the more evolved MS stars should show the biggest decrease in rotation.  Observationally only $V \sin i$ (projected rotational velocity) is determined in these studies due to the orientation of the polar axis to the line of sight ($i$).   But with a large enough sample we can assume a random distribution of this inclination angle.  While it is possible that some massive and evolved objects are observed pole-on, with a large sample it is unlikely that orientation effects can explain the observed trends.

Indirect support that massive stars at low metallcity do NOT slow down comes from a study by \citet{Mae99} who found a significant increase in the number ratio of Be/(B+Be) for decreasing metallicities in Galactic and MC clusters.  They argue that the Be phenomenon is closely linked to rapid rotation, and so if the massive stars with low $Z$ do not slow down, perhaps they produce more Be stars.  We should note here that a Be star is a non-supergiant B-type star that has a significantly lower initial mass than the O-type objects.  The models produced for low $Z$ massive stars do not predict that stars with initial masses relating to Be-type will have metallicity dependent decrease in rotation rates while on the MS.  However the initial velocities of these stars may vary with  $Z$.  The initial rotation rates of massive stars at low $Z$ are not known.  \citet{Mae01} have assumed that the initial rates are the same in the MC and Galaxy (since observational data are lacking).  But the star formation process itself may be altered at smaller metallicity.  Although no numerical models of the formation of massive stars at low $Z$ have been produced, there are some possible origins of higher initial rotational velocities.  A lower $Z$ environment would have lower dust content and fewer metallic ions present in star forming regions so that the magnetic field of the contracting central mass is less coupled to the surrounding region.  Also the lifetime of the accretion disk is likely related to its opacity, which will decrease with lower metallicity.  These factors would result in less angular momentum loss during star formation, leading to higher initial rotational velocities.

There are few direct measurements of the rotation rates of massive stars at low metallicities.  \citet{Mae99} state ``Of course, direct observations of $V \sin i$ in LMC and SMC clusters are very much needed in order to substantiate the above results."  Only a few studies have been published to date on the rotation rates of massive stars at low metallicity.  Some have focused on early B type stars \citep{Kel01,Duf06}, but again we caution that these cooler objects are not exclusively produced by stars with 15 or more solar masses and are often significantly larger than their terminal age main sequence (TAMS) radii.

We made a study of 44 O-type stars in the SMC and LMC utilizing archival HST observations \citep{Pen04}.   We compared the cumulative distribution function (CDF) of the $V \sin i$ values for Galactic, LMC and SMC evolved (giant, bright giant, and supergiant) stars.  The CDF gives the fraction of stars having a $V \sin i$ less than a specific upper value that ranges from zero to the maximum $V \sin i$ observed. The surprising result is that the distributions are very similar, with a slightly larger fraction of LMC stars rotating faster, but a slightly smaller fraction of SMC stars, with lowest $Z$, rotating slower than the Galactic stars.  The two sample Kolmolgorov-Smirnov (K-S) test compares two independent samples and determines a statistic, $p$ describing the similarity of the CDFs.  For low values of $p$ ($\le 5 - 10\%$), the null hypothesis that the two samples are drawn from the same parent population is rejected.  Comparing the LMC and SMC samples with Galactic stars, their K-S $p$ are $44\%$ and $42\%$, respectively, indicating that they are drawn from the same parent population as the Galactic sample. Thus, there is no visible trend of CDFs with metallicity.  The sample of  unevolved (dwarf and subgiant) stars was too small to make a reliable comparison, but the average projected rotational velocities in each group appear to be comparable: $<V \sin i> = 131 \pm 93, 117 \pm 31$, and $78 \pm 45$ km s$^{-1}$ for the Galactic, LMC, and SMC unevolved stars, respectively. The results of this study should be treated cautiously for two reasons.  First, the samples of LMC and SMC targets are not large enough to assume a completely random distribution of inclination angle on the sky.  Second, 18 of the HST/STIS SMC observations were obtained through one observing program.  This program's goal was to study the UV characteristics of massive stars at low Z.  These targets may have a bias toward small projected rotational velocities, in order to better study the line depths as a function of metallicity.

\citet{Hil03} and \citet{Bou03} examined a total of 8 O-type stars in the SMC by fitting rotationally broadened synthetic spectra from the atmosphere model code TLUSTY \citep{Hub95} to the observed spectra.  In all but one case the determined projected rotational velocities were $\le 100$ km s$^{-1}$, much smaller than the predicted equatorial rotation speeds of $\approx 200$ km s$^{-1}$.  There are several possible explanations: (1) the initial rotation rates may not be as large as 300 km s$^{-1}$ for massive stars in the SMC;  (2) the stars may experience more spin down than predicted;  and (3) with only 8 targets, the sample is just too small considering the random distribution of inclination.

The most recent survey of rotational velocities of O-type stars at low $Z$ is a study by \citet{Mok06}, which is part of the larger VLT-FLAMES survey of massive stars \citep{Eva05}.  They determined $V \sin i$ values for 28 O-type stars in the SMC.  A comparison of their unevolved targets (classes V, IV) to evolved targets (classes I, II, \& III) shows that the unevolved sample contains relatively more fast rotators than the evolved stars.  They point out that this trend is also seen in Galactic O type stars \citep{Pen96,How97}.  New stellar interior models predict this behavior for solar metallicity stars but not for massive stars at $Z_{SMC} = 0.004$.  
They also compare their 21 unevolved SMC stars to 66 unevolved Galactic stars from \citet{Pen96}.  The K-S $p$ is only 13\%, suggesting that there is a possible difference in the initial rotation rates of massive stars due to metallicity.  However, \citet{Mok06} warn that because of their small sample size, this result should be verified with a larger SMC sample.

All of the previous $V \sin i$ surveys of massive stars at low $Z$ suffer from one common major fault, a small sample size. Here we present the results of our project to test the treatment of angular momentum in the new stellar interior models through a large scale survey of projected rotational velocities of O-type stars in three metallicity environments: the Milky Way ($Z_{MW} = 0.020$), LMC ($Z_{LMC} = 0.007$), and SMC ($Z_{SMC} = 0.004$). The Far Ultraviolet Spectrographic Explorer (FUSE) archive at the Multimission Archive at Space Telescope (MAST) contains spectra of 161 LMC and SMC stars with spectral classes between B2 - O2.  The targets are 120 evolved (luminosity classes I, II, \& III) and 41 unevolved (IV \& V luminosity classes) stars in these low $Z$ environments.  In addition MAST contains spectra of approximately 100 Galactic O-type stars with known $V \sin i$ values.  This is a large sample of O-type spectra, all observed with the same instrument.  It contains both cluster and field stars.  The observations were not part of any single campaign, in fact they were obtained through 38 different observing programs.  These stars represent a completely uniform, unbiased sample.  %Although only $V \sin i$ can be determined, due to the large sample sizes this is acceptable.  We make the following statistical argument. The inclination angles of the polar axes, $i$, of a group of stars should be randomly distributed. For each metallicity environment (Galactic, LMC, and SMC) we expect that the distribution of $i$ should be the same, that is, random. Thus, we can safely intercompare the rotational properties through an examination of their projected rotational velocities.  
Comparisons of the $V \sin i$ distributions from the unevolved MC stars to similar classed Galactic stars will determine whether the initial rotation rates of massive stars have a metallicity dependence.  Comparisons of the evolved stars in each environment to their unevolved counterparts will show whether the rotation rates are age dependent at that metallicity.  %These are important tests that will constrain the treatment of angular momentum in the new stellar interior models. 

\section{Projected Rotational Velocities from UV lines} 

For the extremely massive O-type stars, the major mechanisms that affect stellar line widths are rotation and macroturbulence.  \citet{Con77} completed the first large-scale survey of rotational velocities of Galactic O-type stars.  They noted in their sample that no O-type star displayed extremely sharp lines that would be expected with a pole-on orientation and suggested that large-scale atmospheric motions contribute to the broadening of the spectral features.  In determining their $V \sin i$ estimates they adopted the convention that only rotational broadening was present.  They were aware this is not in general true and note, as do others who have adopted a similar convention \citep{Pen96,How97}, that their quoted $V \sin i$ is actually a parameter which is closely related to, but not necessarily, the exact projected equatorial velocities of the stars.  Authors who have investigated the amount of macroturbulence present in early B-type supergiants have found values on the order of 20 --70 km s$^{-1}$ \citep{Rya02,Duf06}.  These values do not appear to be dependent upon metallicity as both the Galactic \citep{Rya02} and the SMC \citep{Duf06} samples result in similar values for macroturbulent broadening.  While macroturbulence is undoubtedly a factor in the line widths of O-type stars, we can make a statistical argument that if each sample (Galactic, SMC, LMC) has the same contribution to line widths from macroturbulence, then the distribution of $V \sin i$ from each can be compared for differences in equatorial rotational velocities (see discussion in \S5).  

Our previous studies of projected rotational velocities \citep{Pen96,Pen04} utilized archival high-resolution spectra from the International Ultraviolet Explorer (IUE) and the Space Telescope Imaging Spectrograph (STIS) in the wavelength region 1200 -- 1900 \AA.  The ultraviolet is actually a preferable region for determining projected rotational velocities for very massive stars.  Unlike the photospheric lines seen in optical spectra, the UV lines are formed deeper in the star and are usually less contaminated by wind effects.  Also there are thousands of lines present in the UV, compared to tens in the optical.  The methodology used to obtain $V \sin i$ values from UV spectra differs from that used for optical spectra.    We cross-correlate the spectrum of a narrow-lined star with that of a test star, which results in a cross-correlation function (CCF) that represents a ``superline'' of the test star.  The superline will have an observed width related to the line width of the test star and that of the narrow-lined star.  In the case of a double-lined binary, the CCF may display two peaks, which are separated by the radial velocity difference of the two stars.  A calibration between CCF width and $V \sin i$ is obtained by cross-correlating a sample of stars with known projected rotational velocities.

\section{Observations and Reductions}

The CCF method can also be utilized with FUSE observations of O-type stars. We obtained FUSE archival spectra of 97 Galactic, 55 SMC, and 106 LMC massive stars that are available from MAST.   For each star we obtain the calibrated (CALFUSE) exposure files from the LiF2 channel and detector segment A (hereafter LiF2A spectra). The LiF2A spectra cover only the wavelength region of 1086 -- 1183 \AA.  Below this region the spectra are severely affected by interstellar features and are not suitable for our purpose.   Each observation is actually composed of several exposures.  The exposure spectra were extracted using the CALFUSE routine MRDFITS.  Following this basic reduction we use a series of routines written in the Interactive Data Language\footnote{IDL is a registered trademark of ITT Corporation} to further reduce the spectra. The extracted exposure spectra then were coadded with appropriate wavelength shifts, determined by cross-correlating the exposure with the highest flux value (reference spectrum) to all others, creating a weighted average spectrum based upon exposure lengths for each observation.  These steps are consolidated into one routine COMBINE that writes the coadded spectrum to a file.   Data spikes are removed in SPIKEOUT, which replaces any bad pixels with a median of surrounding pixels.  One of the most important steps in this reduction is the placement of the data on a log $\lambda$ grid, in which each pixel step corresponds to a uniform velocity step of 1 pixel = 10 km s$^{-1}$.  This is done in the routine BINFUSE.  The spectroscopic resolving power of the raw FUSE spectra is 20,000 $\pm$ 2000, however we further degrade this value by smoothing the data by a Gaussian transfer function (FWHM = 40 km s$^{-1}$) in GSMOOTH.  Another important step is the alignment of all spectra of a star on the same wavelength system.  In order to assure that all spectra of a given star are not shifted with respect to each other, the interstellar lines present (assumed to have a constant velocity) are used to align the system spectra.  ISMALIGNFUSE does this by creating an average spectrum, cross-correlating the regions containing strong interstellar lines (which are picked interactively), and shifting each spectrum to an average wavelength grid.  ISMALIGNFUSE also removes not only the strong interstellar features, but also those chosen (again, interactively) to be weak interstellar features.  We cannot overemphasize the importance of removing the considerable number of interstellar lines, whose identification was greatly aided by earlier work of \citet{Bar00}, \citet{Pel02}, and \citet{Wal02}.  The spectra are then rectified (routine FUSERECT) to a unit continuum, by fitting a spline function to a series of pseudo-continuum zones in an average spectrum and then dividing each of the spectra by the fit.  The final spectral stacks are written to a file in WTSTKFUSE.  All of these procedures are incorporated in a calling routine named MKSTKFUSE. 

\section{Projected Rotational Velocities from Cross-Correlation Functions}

Target star spectra are cross-correlated with that of a relatively narrow-lined template star.   Cross-correlation was done only in the wavelength regions 1100 -- 1108 \AA ~and 1130 - 1165 \AA ~(1139 - 1165 \AA, for stars with emission in the \ion{P}{5}/ \ion{Si}{4} $\lambda 1128$ line), a region relatively free of wind features.  While most of the lines in these regions are too weak and blended to identify individually, a few of the stronger features are \ion{O}{3} $\lambda\lambda 1149.6, 1150.9, 1153.8$ \AA, and \ion{S}{4} $\lambda 1138.2$ \AA. The spectra are padded with $1000$ km~s$^{-1}$ of artificial continuum on both ends to include the entire observed range in the cross-correlation.  The CCF is the sum of the square of the differences between the test spectrum and the reference spectrum shifted in velocity from $-1000$ km~s$^{-1}$ to $+1000$ km~s$^{-1}$ at $10$ km~s$^{-1}$ intervals.  The functions are then rectified, inverted, and Gaussian fitted to obtain an estimate of the full-width at half-maximum (FWHM).  In Figure 2 we present fitted CCFs for 4 Galactic targets, HD 46202, HD 152233, HD 96917, and HD~41161 that were all made with the template spectrum of star AV~327 (O9.7 II-Ibw).   We note that the `bumps' present in the CCF of HD~41161 are typical for massive stars with large projected rotational velocities and are possible indicators of non-radial pulsations present in the star.  

The next step is developing a calibration of CCF width to $V \sin i$ using the Galactic targets.  These stars have known $V \sin i$ values from \citet{How97}.  We created calibrations for two template stars AV 327 (O9.5 II-Ibw) and SK -66 100 (O6 II(f)) in order to cover the large range in spectral class of our targets.  These are presented in Figure 3.  Because of the early type of SK -66 100, seven of the latest type stars did not produce significant CCF peaks and were not used to create the calibration for this template.  In Table 1, we list the Galactic targets along with their spectral classes, $V \sin i$ values from \citet{How97}, $V \sin i$ values computed from the calibrations of each template star, and the FUSE observations used.  Spectral classes are as listed in \citet{How97}.

For each target from the Magellanic Clouds, we follow a similar procedure.  All targets are cross-correlated with both templates to determine which template creates the CCF with the highest peak.  For target stars with more than one spectrum, each spectrum's CCF is examined to determine if the peak is stationary.  A CCF peak that moves from observation to observation is indicative of a single-lined spectroscopic binary, as in the case of NGC346-MPG324 which shows a peak shift of  $110$ km~s$^{-1}$ between two observations.   In these cases, each CCF is fit individually.   For targets with more than one observation and stationary peaks, the average CCF is fitted.  A few of the stars' CCFs had double peaks, indicating that they were double-lined spectroscopic binaries.  These CCFs were fit with a double Gaussian.  Then using the calibrations developed above a $V \sin i$ value is determined from the Gaussian width of each fitted CCF.  In Tables 2 and 3 we list the SMC and LMC targets, their spectral classifications, $V \sin i$ values from this paper, along with any previous values, the template star utilized, FUSE observations, and an additional column for other comments.  For those CCF with two peaks the weaker peak's $V \sin i$ is given in parentheses.  In Figure 4 we plot our $V \sin i$ values versus those from \citet{Mok06,Mok07}  and \citet{Pen04} for those targets in common.   From the scatter in Figure 4, we estimate that the typical errors for our $V \sin i$ values are 10 km s$^{-1}$.

\section{Results and Conclusions} 

We can compare the projected rotational velocity distributions of the 
different groups of stars by calculating their cumulative distribution function (CDF).  Again, the CDF gives the fraction of stars having a 
$V \sin i$ less than a specific upper value that ranges from zero to the maximum $V \sin i$ observed.  Not all of our targets stars are used in compiling the CDF.   Stars with a `NO' in the final column of Tables 2 and 3 are excluded from our analysis for various reasons.  For the B-type stars in our sample, we perform an additional analysis to determine: if they formed from stars with initial masses larger than 15 $M_{\odot}$, and the fraction of their current radii to their TAMS radii.   For all the stars we also wish to insure that extremely close binary systems are not present.  We calculate $M_{bol}$ from their $V$, $B-V$ values, the $(B-V)_0$ calibration of \citet{Shu85}, $A_V = 3.1\times E(B-V)$, bolometric corrections of \citet{How89}, and a distance modulus of 18.3 and 19.1, for the LMC and SMC respectively.  We adopt the effective temperatures from the calibration of \citet{How89} for O-types and \citet{Boh81} for B-types.   From the evolutionary tracks of \citet{Sch93}, we extrapolate initial masses, current masses, log $g$, current radii, and TAMS radii.  As we expected a large number of our B-type targets are inappropriate for our analysis.   None of the B-stars have initial masses smaller than $15 M_{\odot}$.  However, thirty-six targets (11 SMC and 25 LMC) are a factor of 50\% or more larger than their TAMS radii.  The lack of lower mass B-type supergiants almost certainly reflects their lower luminosities and the subsequent difficulty of obtaining FUSE spectra of those stars.  For the entire FUSE sample we also compare the interpolated log $g$ values with the luminosity class for discrepancies which might indicate that more than one star is present in the spectrum.  For our analysis we also include 27 SMC and 26 LMC stars from \citet{Mok06,Mok07} and \citet{Pen04}.  We have already shown in Figure 4 that there are no systematic differences between the $V \sin i$ values from our FUSE analysis and those presented in these prior studies.  These additional stars are presented in Tables 4 and 5, along with the spectral classification given in \citet{Mok06,Mok07} or \citet{Pen04}, and those authors' $V \sin i$ values.

We divide the data into the following samples: SMC unevolved (32 stars), SMC evolved (19 stars), LMC unevolved (36 stars), and LMC evolved (42 stars).  We also create samples for Galactic unevolved (79 stars) and evolved (56 stars) from \citet{How97}.  Following the convention of earlier studies \citep{Pen96,How97,Pen04,Mok06,Mok07}, unevolved refers to stars with luminosity classes V \& IV, while luminosity classes I \& II are termed evolved.  Class III stars are omitted from both groups.  Three stars are without explicit luminosity classes, AV~80, SK -67 266, and SK -69 279, and we assign these to the evolved category based on the notes from their listed spectral classification source.  For each sample we create a CDF and plot these in comparison with other samples to determine the similarity.   For each comparison a K-S statistic, $D$, and its corresponding probability, $p$, are calculated and these are presented in Table 6.   The test statistic, $D$, is the maximum vertical variance between the two CDFs at any $V \sin i$ value.  The $p$ value tells us the probability of obtaining a $D$ value at least as extreme as was observed, assuming that the two samples are drawn from the same parent population.   So in cases where the $p$ statistic is small, i.e., $\le 5\%$, we can say that the probability of finding such a large $D$ between two samples that are drawn from identical populations is very small.  Plots of the comparisons made are presented in Figures 5 -- 7 and 9 -- 10.  At the suggestion of the referee, we determined the mass distribution of each of our samples.  For the MC stars, we adopted the current evolutionary masses extrapolated from the $M_{bol}$ and $T_{\rm eff}$ analysis described in the paragraph above.  Mass estimates for the Galactic sample can be found in \citet{How89}.   The mean masses of the samples are 37.8, 45.7, and $34.4 M_\odot$ for the unevolved stars of the Galaxy, LMC, and SMC, respectively, and the means are systematically
higher among the evolved stars of the same three environments,
45.4, 47.4, and $40.3 M_\odot$, respectively.  This difference is
expected since the evolved O-star samples will be biased towards
higher masses (the advanced evolutionary stages of lower mass stars
occur at cooler temperatures).  The K-S test indicates that the three
evolved sets have statistically indistinguishable mass distributions.
On the other hand, the K-S test suggests that some differences may
exist among the unevolved star mass distributions (particularly
between the SMC and LMC samples where $p=0.7\%$), but given the
relative small sample sizes, we doubt that these are significant.

First we examine the evolved and unevolved Galactic samples to determine the level of difference we would expect in an environment where the stars do slow down as they evolve (Fig.~5).  We see that the maximum variance in their CDF is 0.30 and this occurs at $V \sin i$ of $83$ km~s$^{-1}$.  This large a $D$ value results in a $p$ of $0.5\%$, but we should not be misled by this low value.  At these smaller velocities the source of the difference between the evolved and unevolved samples is not a result of angular momentum loss, but from the larger amount of macroturbulence that is present in the evolved stars' photospheres, broadening their line profiles.  At the higher velocities where we expect to see evidence of spin down, the largest divergence is $\approx 0.12$, which would give us a significantly higher $p$ of $71\%$.  This is not say that the evolved stars in the Galactic sample have not slowed down, but that this effect may be more subtle than we originally expected.   

In the LMC, the same comparison of unevolved to evolved stars has a very different appearance (Fig.~6).  The CDF of both samples appear very similar, and the derived $D = 0.15$ and corresponding $p = 75\%$ supports the null hypothesis that both are drawn from the same parent distribution.  This result is in agreement with the suggestion of \citet{Wol08}, using data from \citet{Hun08}, that massive stars in the LMC have similar $V \sin i$ distributions regardless of their $\log g$ values.   Examining Figure 6, we notice that there is almost no divergence at low $V \sin i$ values, and a smaller variance at larger $V \sin i$ than we see in the Milky Way samples.   The very high $p$ value here is primarily due to the presence of slow rotators among both the unevolved and evolved samples, unlike the Galactic evolved group.  At high $V \sin i$ values, the CDFs of evolved to unevolved differ by $\approx 0.09$ which is only slightly smaller than the value for the Galactic samples.  It is at the low velocity end of the CDF where the LMC samples differ from those of the Galactic samples. The good agreement between the low ends of the two CDFs for the LMC stars suggests that development of macroturbulence with evolution is not as large a factor in the photospheres of the massive stars in that environment and that the processes that lead to macroturbulent broadening may have a metallicity dependence.  

The CDFs of the same populations in the SMC are presented in Figure 7.   Although the $p$ value from the KS test is $23\%$,  which would indicate that the samples are drawn from the same parent population, the maximum difference $D$ is $0.29$, which is very close to that from the Galactic samples.  The larger $p$ value from the similar $D$ reflects the much smaller sample sizes in the SMC.   At the high velocity end of the CDF, the fraction of evolved stars with $V \sin i$ above 200 km s$^{-1}$ is $11\%$ compared to $13\%$ for the unevolved.  This does support the hypothesis that stars at SMC metallicity will not slow down as they evolve on the MS.  At the other end of the CDF, there is a slightly larger fraction of unevolved stars with $V \sin i$ below $\approx 70$ km s$^{-1}$, but this difference is much smaller than we see in the Galactic comparison.  The real divergence between the evolved and unevolved CDFs comes at the intervening velocities.  The maximum divergence, 0.29, occurs at $V \sin i = 107$ km s$^{-1}$.  The fractions of stars with $V \sin i$ at or below this value for the evolved and unevolved samples are 0.84 and 0.55, respectively.   This trend was also seen by \citet{Mok06}, who surmised that the initial rotational velocity distribution in the SMC might vary significantly from that in the Galaxy.   Why we see a disparity in the CDFs of the evolved and unevolved samples in between $80 - 190$ km s$^{-1}$ only in the SMC samples is not clear.  We emphasize that the K-S test results accept the null hypothesis that the unevolved and evolved samples in the SMC are drawn from the same parent distribution. 

A primary purpose of this project is to examine observationally the amount of angular momentum loss during the MS lifetimes of massive stars and the effects of metallicity on this loss.  Comparing the evolved to unevolved samples in Figures 5-7, we see that the $D$ statistic at high $V \sin i$ values range from 0.12, 0.09, 0.02 in the Galaxy, LMC, and SMC, respectively.   Taken by itself this is suggestive of a trend with decreasing $Z$.  However, for the LMC and SMC comparisons, $D$ statistics of $0.09$ and $0.02$ result in $p$ values of $99.6\%$ and $100\%$, indicating there is no statistical difference between the evolved and unevolved CDFs at high $V \sin i$.  But this is also true in the Galaxy where we do expect to see the significant spin down between the evolved and unevolved samples.  Here the corresponding $p$ for a $D$ of $0.12$ is $71\%$, far above the $5\%$ cutoff.  Statistically the loss of angular momentum for these H-burning stars is not so different between these three metallicity environments.  

Although our results may appear to cast doubt on the
rotational evolution predicted by models, we suspect that
the rather small changes we find are probably consistent
with models.  Unfortunately, at the moment there are no
large grids available for rotational evolution that
might be used to compare with our statistical results.
The best set was calculated by \citet{eks08} who present
evolutionary tracks for model 20 and $60 M_\odot$ stars.
The tracks for the $20 M_\odot$ models are not immediately
applicable since their portions that correspond to
luminosity classes I and II occur in the cooler B-star
domain (not included in our sample statistics).
On the other hand, the tracks for $60 M_\odot$ stars
probably only apply to a small fraction of our sample
of stars.  However, the $60 M_\odot$ tracks do cover
all luminosity classes, so we can use them to make a
representative comparison of the predicted and observed
changes in rotational velocity with age.  We plot in
Figure~8 the $60 M_\odot$ models from \citet{eks08}
that show the change in the average $V\sin i$ with age.
Here we have multiplied the model equatorial velocities
by a factor of $\pi / 4$ to represent the average
projected velocity for random orientations.
The thin solid lines show the velocity evolution for
Galactic abundance stars for three values of the initial
ratio of actual to critical angular velocity, and
the thin dotted lines show the same for lower metallicity
SMC stars.  Above these are three tick marks indicating
those ages associated with luminosities and temperatures
of luminosity class V, III, and I stars according to the
observational calibration of \citet{mar05}.  For simplicity,
we will assume that our sample groups of stars correspond
to ages midway between 1 Myr (approximately when massive stars
emerge from their natal, ultracompact H~II region) and
1.83 Myr (the class III age) for the unevolved group
and midway between 1.83 Myr and the age at which they cool to
$T_{\rm eff} = 28$~kK (the boundary between O- and B-supergiants)
for the evolved group.  We plot our derived average $V\sin i$
values for these groups (Table 6) versus their assigned
mid-ages (relative to the $60 M_\odot$ tracks)
as thick solid and dotted lines bounded by plus signs
for the Galactic and SMC samples, respectively.
We see that the modest declines observed are not too different
from the model estimates for stars with relatively small rotational
velocities.  The tracks for lower mass stars
are qualitatively similar, but their later, evolved portions where the
rotational velocities change the most will be progressively shifted out
of the O-star domain into cooler classes, so that the predicted
changes in mean rotational velocity between unevolved
and evolved samples will be even smaller than shown in Figure~8.
Thus, we emphasize that our simple comparison is probably only valid for considering relative trends with age and metallicity, and a detailed comparison would require additional
tracks for other masses and sampling of the tracks based upon the
observed mass and $V\sin i$ distributions.

We also are interested in whether the initial $V \sin i$ distribution is the same in the three environments.  In Figure 9, we plot the CDFs of all three dwarf samples.  Statistically all three have $p$ values that are above the cut-off of $5\%$, and certainly at the high $V \sin i$ end all three look extremely similar, indicating that the maximum rotational velocities are very similar.  Below $200$  km s$^{-1}$, the behavior of the three distributions varies.  Again we see that the shape of the CDF for the SMC dwarfs is dissimilar from that of the counterparts in the Galaxy  and  the LMC between $80 - 190$ km s$^{-1}$.  Below $80$ km s$^{-1}$, the fraction of slowly rotating stars varies between the three environments, with the Galactic sample in between the SMC and LMC.  In fact the largest divergence between the LMC and SMC CDFs comes at $V \sin i \approx 65$ km s$^{-1}$, resulting in a $p$ value just above the statistically significant level.   \citet{Hun08} showed an analogous plot, but for objects with $M < 25 M_{\odot}$.  Similarly they find good agreement between the LMC and Galactic samples, with the SMC sample lying slightly beneath the other two, especially in the region below $200$ km s$^{-1}$.  It is interesting that there are no very slow rotators in our LMC dwarf sample.  At these low velocities, the dominating effect must not depend upon metallicity since the Galaxy's metallicity, with significantly higher $Z$, is situated between that of the two low $Z$ samples.  We conclude that the initial velocity distribution in our three unevolved samples are statistically indistinguishable.  

%This is a somewhat different conclusion than that reached by \citet{Hun08} with regard to objects with $M < 25 M_{\odot}$, who state that the SMC objects are rotating faster than their Galactic counterparts at the $3\sigma$ level.  However in examining their plot, at high velocities all three CDF appear very similar.  For both their and our own results, the 

Finally we compare the evolved stars in each environment to examine the relative fractions of stars with large $V \sin i$ (Fig.~10).  Looking at the distributions at the high velocity end, we see that a slightly larger fraction of evolved stars in the SMC have $V \sin i$ values larger than $200$ km s$^{-1}$ than in the Galaxy or LMC, which might support theoretical predictions.  But the large $D$ values that are found between the LMC and SMC samples are again at the low velocity range, $\approx 80$ km s$^{-1}$.  The maximum differences between the Galaxy and both the LMC and SMC CDFs occur near this velocity, with both of the low metallicity environments having a larger fraction of stars with smaller values compared to the Milky Way sample.  In fact the trend is very supportive of our earlier suggestion that metallicity plays an important role in macroturbulent broadening in evolved O-type stars.  A recent paper by \citet{Can08} discusses the origin of atmospheric turbulence in massive stars by sub-surface convection zones that are driven by Fe-peak element ionizations.  In their simulations, the threshold luminosity for the occurrence of an iron convective zone is ten times lower at $Z_{MW}$ than that for $Z_{SMC}$.  Our results support their prediction that turbulence will increase with metallicity.  The SMC evolved sample has the largest fraction of slow rotators, followed by the LMC and then the Galaxy.  We note that the $D$ values for the SMC vs. Galaxy and LMC vs. Galaxy are extremely similar, but result in differing $p$ values owing to the smaller SMC sample size.  The large divergence at small $V \sin i$ values between the Galaxy and LMC, and possibly SMC,  leads us to reject the hypothesis that they are both from the same parent distribution.   Again we stress that the differences that cause this are not the fractions of  stars with large $V \sin i$, but those with small values.  

In conclusion, we find some support for the new stellar interior model predictions that massive stars in lower metallicity environments will remain at almost constant rotation rates throughout their MS lifetimes.  But we also see that this effect, loss of angular momentum during the MS, is more subtle than previously reported even at the higher Galactic metallicity.  We find that the ratio of mean $V\sin i$ in the evolved to unevolved samples is 0.84, 0.90, and 0.88 for the O-stars of the Galaxy, LMC,
and SMC, respectively. This is consistent with the general expectation of spin down and the specific expectation that the spin down rates are large in the Galaxy where mass loss rates are higher.  Note that the ratio of 0.84 for the Galaxy is an upper limit, since the apparent
projected velocities of the Galactic evolved stars are
systematically higher because of larger macroturbulence.  With the exception of the SMC unevolved sample, the largest divergence between sample CDFs occur at low $V \sin i$ values.   We have suggested that metallicity may play an important role in the development of macroturbulence in the photospheres of the evolved, massive stars.  Finally, we once again see that small sample sizes, particularly for the SMC, hamper our ability to establish the true distribution of equatorial rotational velocities.  We urge the continued observational efforts to determine the $V \sin i$ values for a large number of massive stars in differing $Z$ environments to address this pressing and important issue.

\acknowledgments 

The FUSE data presented in this paper were obtained from the Multimission Archive at the Space Telescope Science Institute (MAST).  STScI is operated by the Association of Universities for Research in Astronomy, Inc., under NASA contract NAS 5-26555.  Support for MAST for non-HST data is provided by the NASA Office of Space Science via grant NAG 5-7584 and by other grants and contracts.  This research has made use of the SIMBAD database, operated at CDS, Strasbourg, France.   Support for this work was provided by program GO63 from FUSE and  NASA 06-ADP06-68.
Institutional support for L.R.P. has been provided from the College of Charleston School of Sciences and Mathematics.  Additional support for L.R.P.\ was provided from AR Nos. 09945 and 11275 from STScI,   NASA grant NAG 5-2979, and NSF grant AST-0506541.  Additional support for D.R.G. was provided by the National Science Foundation under Grant No.~AST-0606861.  We gratefully acknowledge all this support.

\appendix
\begin{center}
APPENDIX:  NOTES ON INDIVIDUAL STARS
\end{center}

\noindent{AV~80 -- This star is included in the CDF for SMC evolved stars in accord with the individual notes in \citet{Wal00}, who state that its absolute visual magnitude and the strength of its UV wind features correspond to a luminosity class of III.}

\noindent{AV~177 -- The CCF has a double-peaked appearance, however the velocity separation is not large enough for a double Gaussian fit.  The $V \sin i$ value listed is of both peaks together.}

\noindent{AV~332 -- Known as a double-lined spectroscopic binary classified as WN3 + O6.5 I \citep{Mas00}.  Only one peak is seen in the CCFs, probably that of the O6.5 I component, which moves in $V_r$ by 45 km s$^{-1}$ between the observations taken 6 months apart.}

\noindent{AV~476 -- Classified as O2-3 V + companion by \citet{Mas05}, we see two peaks, separated by $V_r$ of 363 km s$^{-1}$.  

\noindent{NGC346-MPG324 -- The peak in the two observations moves by 110 km s$^{-1}$ and changes width.}

\noindent{SK -65 21 -- There are four observations taken in sequence over 15 hours.  All CCFs show double peaks that appear to be near or at maximum velocity separation of 350 km s$^{-1}$.  The peaks first move slightly farther apart and then closer over the four observations.}

\noindent{SK -66 172 -- Classified as O2 III(f*) + OB \citep{Wal02}, we do not see any sign of a companion in the CCF.}

\noindent{SK -67 18 -- \citet{Koe03} classified this system as O3 + O8-B0 I.  We see both peaks in the CCF.}

\noindent{SK -67 166 -- Observed 63 times with FUSE over approximately a 5 week period, the peak does not move. }

\noindent{SK -68 52 -- A weak second peak is stronger in the CCF with AV327 than with SK -66 100.  The peaks are separated by 350 km s$^{-1}$.}

\noindent{SK -69 94 -- The FUSE spectrum of S Dor is contaminated by the visual companion 13$\arcsec$ away which is classified as a B1 I \citep{Wol80}.  The CCF here is almost certainly representative of this star, not S Dor.  

\noindent{SK -69 223 -- Classified WC4(+OB)  by \citet{Bar01}, we do not see any sign of a second peak in the CCFs.  However the main peak does move slightly between the two available observations taken 50 minutes apart.} 

\noindent{SK -70 91 -- Classified a O2 III(f*) + OB by \citet{Wal02}, both stars have peaks in CCF.  The peaks are separated by 255 km s$^{-1}$.}

\noindent{BI 173 -- In the two observations taken two months apart, the main peak moves by 53 km s$^{-1}$.   There is a low level second peak, but it is difficult to distinguish it from the surrounding noise.}

\noindent{HV~2241 -- A known binary, we do not see any sign of the secondary, classified as early B \citep{Mok06}, in the CCF.}

\noindent{LMC X-4 -- This spectrum is of the O8 III companion in this X-ray binary.}

\noindent{PGMW 3209 -- Only one extremely noisy observation is available from FUSE.  We do not see any sign of a second peak in the CCF.}

%%%%%%%%%%%%%%%%%%%%%%%%%%%%%%%%%%%%%%%%%%%%%%%%%%%%%%%%%%%%%%%%%%%%%%%%%%%
% References

%\bibliography{mybib}{}
%\bibliographystyle{plain}

\clearpage

%%%%%%%%%%%%%%%%%%%%%%%%%%%%%%%%%%%%%%%%%%%%%%%%%%%%%%%%%%%%%%%%%%%%%%%%%%%
 % Table 1 - V sin i for Galactic stars
\begin{deluxetable}{lccccl}
%\rotate
\label{t1}
\tablewidth{0pt}
\tablecaption{Galactic Targets}
\tabletypesize{\scriptsize}
\tablecolumns{6}
\tablehead{ 
\colhead{}  &  \colhead{} &  \multispan{3}\hfil {$<V \sin i>$}\hfil  & \colhead{} \\ 
\colhead{} &\colhead{Spectral}  &    \colhead{\citet{How97}}  &
 \colhead{AV 327}  & \colhead{SK -66 100} & \colhead{FUSE} \\
\colhead{Star} & \colhead{Classification} & \colhead{(km~s$^{-1}$)}& \colhead{(km~s$^{-1}$)} 
& \colhead{(km~s$^{-1}$)} 
 &\colhead{data set} 
}
\startdata 
HD~108    &  O6:f?pe &    78 &    66  & 82 & O63508010 \\ 
HD~5005A	& O6.5 V((f))	& 99 &	78&		108&	P1020102\\
HD~12323&	ON9 V	&131 &	99&	73	&P1020202\\
HD~13268&	ON8 V &	309&	264&	252&	P1020304\\
HD~14134&	B3 Ia&	66&	85&	\nodata	&D9021001\\
HD~14434&	O5.5 Vn((f))p&	423&	412&		436&	P1020504\\
HD~15642&	O9.5 III:n&		331&	398&		\nodata&	M1122301, P1020702\\
HD~34078	&	O9.5 V&	30&	45&		54&	A0700203, B0630102, C0730101-2, E0760101\\
HD~34656&	O7 II(f)&	91&	70&		75&	P1011301\\
HD~41161&	O8 Vn&	304&	314&		353&	P1021001\\
HD~42088&	O6.5 V&	65&	68&		66&	P1021101\\
%HD~43384	&	B3 Iab&	59&	121&		\nodata&	P2160401\\
HD~45314&	OBe &	285&	308&		227&	P1021301\\
HD~46056&	O8 Vn&	330&	274&		310&	P1160901\\
HD~46149&	O8.5 V&	78&	75&		66&	C1680401\\
HD~46150&	O5 V((f))&	111&	78&		88&	C1680201, P1021401-3\\
HD~46202	&	O9 V&	37&	43&		46&	P1161001\\
HD~46223	&	O4 V((f))&	82 &	68&		114&	C1680302\\
HD~60369&	O9 IV &	67&	74&		62&	P1050201\\
HD~61347&	O9 Ib &	116&	98&		106&	P1022001-2\\
HD~63005&	O6 V((f)) &	74&	72&		98&	P1022101\\
HD~69106&	B0.5 IVnn &	325&	309&		353&	P1022301-2\\
HD~72648&	B2.5 II-III &	48&	45&		\nodata&	A1290401\\
HD~74920&	O7 IIIn &	251&	298&		286&	P1022601\\
HD~89137&	ON9.7 III(n)p &	202&	231&		137&	P1022801\\
HD~90087&	O9.5 III &	272&	305&		259&	P3030401, P1022901\\
HD~91572&	O6 V((f)) &	62&	67&		75&	P3031701\\
HD~91651&	O9 V:n &	292&	318&		294&	P1023101-2\\
HD~91824&	O7 V((n)) &	65&	69&		70&	A1180802\\
HD~92554	&	O9.5 IIn &	298&	348&		308&	P1023201-2\\
HD~92850	&	O9.5 Ib &	78&	61&		53&	P3031601\\
HD~93028&	O9 V	&	42&	65&		49&	A1180902\\
%HD~93129A&	O3 If*&	180&	62&		120&	P1170202\\
HD~93146	&	O6.5 V((f)) 	&	79&	83&	78&	P1023301\\
HD~93222	&	O7 III((f)) &		77&	87&	81&	P1023701\\
HD~93249&	O9 III 	&	78&	69&	71&	Z9013401\\
HD~93250&	O3 V((f)) &	107&	162&		129&	P1023801\\
HD~93843&	O5 III(f) var &	95&	96&		104&	P1024001\\
HD~96670	&	O8 p &	139&	111&		120&	P1024201\\
HD~96715	&	O4 V((f)) &	76&	75&		96&	P1024301\\
HD~96917	&	O8.5 Ib(f) &	176&	178&		178&	P1024401\\
HD~99546&	O8 &	79&	68 &		65&	P3031901\\
HD~100444&	O9 Ib &	83&	72&	75&	Z9014101\\
HD~101008&	O9 IV &	60&	67&	47&	Z9014201\\
HD~101190&	O6 V((f)) &	88&	83&		118&	P1025001\\
%HD~101205&	O7 IIIn((f)) &	186&	296&	189&	P1025101\\
HD~101298&	O6 V((f)) &		80&	79&		95&	P1025201\\
HD~101413&	O8 V	 &	99&	94&		90&	P1025301\\
HD~101436&	O6.5 V &	98&	127&	128&	P1025401\\
HD~102475&	B0.5 II &	56&	65&		53&	G0630301\\
HD~104683&	B1 Ib &	59&	48& \nodata&	Z9014401\\
HD~105056&	ON9.7 Iae var &	68&	70&		58&	Z9014501\\
HD~111973&	B3 Ia &	68&	73&		\nodata&	Z9014501\\
HD~112784&	O9.5 III &	51&	58&		51&	Z9015001\\
HD~113012&	B0.2 Ib &	51&	58&		48&	Z9015101\\
HD~115455&	O7.5 III((f)) &	69&	84&		85&	A1200701\\
HD~116852&	O9 III &	136&	114&		95&	P1013801\\
HD~122879&	B0 Ia &	92&	69&		58&	B0710501\\
HD~124314&	O6 V(n)((f)) &	242&	285&		302&	P1026201\\
HD~124979&	O8 ((f)) &	220&	252&		280&	P1026301\\
HD~148937&	O6.5 f?p &	76&	77&	90&	C1680101\\
HD~152233&	O6 III:(f)p &	104&	102&		102&	P1026702\\
HD~152249&	OC9.5 Iab &	99&	83&		72&	B0250401\\
HD~152314&	O9.5 III-IV	&	76&	86&		72&	P1026901\\
HD~152405&	O9.7 Ib-II	&	77&	68&	56&	P3031801\\
HD~152424&	OC9.7 Ia &		86&	74&		67&	Z9016301\\
HD~152590&	O7.5 V &	60&	76&		76&	B0710601-2\\
HD~152623&	O7 V(n)((f)) &	106&	97&		94&	P1027001\\
HD~152723&	O6.5 III(f) & 	111&	104&		107&	P1027102\\
HD~154368&	O9.5 Iab &	102&	89&		78&	P1161901\\
HD~157857&	O6.5 III(f) &	97&	99&		151&	P1027501\\
HD~163758&	O6.5 Iaf &	90&	84&		83&	P1015901\\
HD~163892&	O9 IV((n))	&	201&	287&		254&	P1027601-2\\
HD~167659&	O7 II(f) &	85&	77&		95&	P1028001\\
HD~168076&	O4 V((f)) &	98&	77&		116&	P1162201\\
HD~168941&	O9.5 II-III &		116&	96&		82&	P1016501-2\\
HD~175754&	O8 II((f)) &	177&	207&		193&	P1016802\\
HD~175876&	O6.5 III(n)(f) &	266&	274&		293&	P1016902\\
HD~186890&	O7.5 III((f)) &	83&	88&	85&	P3030201\\
HD~189957&	O9.5 III &	92&	92&		78&	P3030301\\
HD~190429A&	O4 If+ &	105&	77&		73&	P1028401\\
HD~190864&	O6.5 III(f) &	88&	85&		91&	E0820501\\
HD~191423&	O9 III:n* &	436&	356&	336&	P3030101, E0821301\\
HD~192639&	O7 Ib(f) &	96&	110&	100&	C1710101, P1162401\\
HD~193514&	O7 Ib(f) &	94&	94&	94&	E0820701\\
HD~199579&	O6 V((f)) &	73&	70&	82&	P1162501\\
HD~201345&	ON9 V &	91&	69&	58&	P1223001\\
HD~207198&	O9 Ib-II &	91&	71&	69&	P1162801\\
HD~207538&	O9.5 V &	51&	58&	36&	P1162902-3\\
HD~210809&	O9 Iab &	117&	98&	81&	P1223101-3\\
HD~216532&	O8.5 V((n)) &	189&	225&	123&	A0510202\\
HD~217086&	O7 Vn &	332&	310&	343&	E0820801\\
HD~218915&	O9.5 Iab &	94&	75&	74&	P1018801\\
HD~225094&	B3 Ia 	&	68&	77 &		\nodata&	Z9120101\\
HD~303308&	O3 V((f)) &	111&	85&	137&	P1221601-2\\
BD -48 3437&	B1 Iab &	54&	44&	\nodata&	P1018401-3\\
BD -60 2522&	O6.5 (n)(f)p &	178&	249&	228&	D1130101\\
CPD -59 2600&	O6 V((f)) &	142&	135&	182&	P1221401-3\\
CPD -74 1569&	O9.5 V &	64&	68&	59&	P1015301, U1092701\\
\enddata
\end{deluxetable}

% Table 2 - V sin i for SMC stars
\begin{deluxetable}{llcclcll}
%\rotate
\label{t2}
\tablewidth{0pt}
\tablecaption{SMC Targets}
\tabletypesize{\scriptsize}
\tablecolumns{8}
\tablehead{ 
\colhead{}  &  \colhead{}&  \colhead{}  &  \multispan{2}\hfil {$<V \sin i>$}\hfil  & \colhead{} & \colhead{}& \colhead{} \\ 
\colhead{} &\colhead{Spectral}  &  \colhead{} & \colhead{This paper}  &
 \colhead{Other}  & \colhead{Template} & \colhead{FUSE} & \colhead{} \\
\colhead{Star} & \colhead{Classification} & \colhead{Ref.} & \colhead{(km~s$^{-1}$)} & \colhead{(km~s$^{-1}$)} & \colhead{Star} 
& \colhead{data sets} 
 &\colhead{Notes} 
}
\startdata 
AV 14&	O3-4 V + neb & 1 &	261 &	212\tablenotemark{a}&	AV 327 &	P1175301& \\	 
AV 15&	O6.5 II(f) & 2 &	95&	128\tablenotemark{b} 132\tablenotemark{c} 135\tablenotemark{a}&	AV 327&	P1150101&    \\	 
AV 26&	O7 III+neb& 1 &	141&	127\tablenotemark{b} 128\tablenotemark{a}&	AV 327&	P1176001&   \\	 
AV 47&	O8 III((f))& 2 &	79&	76\tablenotemark{b} 169\tablenotemark{c}&	AV 327&	P1150202&    \\	 
AV 61&	O5 III & 3 &	228&	 &	SK -66 100&	E5110201&  \\	 
AV 69 &	OC7.5 III((f))  & 2 & 	94&	100\tablenotemark{b} 104\tablenotemark{c}&	AV 327	&P1150303& \\	 
AV 70&	O9 Ia & 3 &	62&	 	&AV 327&	A1180202-3& \\	 
AV 75&	O5 III(f+) & 2 &	94&	109\tablenotemark{b} 110\tablenotemark{c}&	AV 327&	P1150404& \\	 
AV 80&	O4-6 n(f)p  &  2 &	278&	324\tablenotemark{b} 270\tablenotemark{c}&	AV 327 &	 E5111701& * \\	 
AV 83 &	O7 Iaf+   &   2 &	71 &	82\tablenotemark{b} 80\tablenotemark{c} 70\tablenotemark{f}&	AV 327&	 P1176201& \\
AV 95 &	O7 III((f))   &   2  &	76 &	82\tablenotemark{b} 89\tablenotemark{c} 68\tablenotemark{a}&	AV 327&	 E0540101, P1150505& \\	 
AV 96 &	B1 Ia   &   3&	70 &  &	 AV 327&	 C1600101& NO\\	 
AV 104&	B0 Ib   &   3&	56 &    & 	AV 327&	 D1260101& \\	 
AV 119&	B2 II   &   1& 	48 &    &	 	AV 327&	 Z9122101& NO \\	 
AV 120&	O9.5 III   &   4&	70 &   &	 	AV 327&	 E5110301&  \\	 
AV 175&	B2 Iab   &   3&	51 &    &	 	AV 327&	 Z9122001& NO \\	 
AV 177&	O4 V   &   3& 	428 &	 	&SK -66 100&	 C0020102&	*, NO\\
AV 207&	O7 V   &   1&	75 &   	&AV 327&	 P1175901& \\	 
AV 210&	B1.5 Ia   &   3&	43 &	37\tablenotemark{d}&	AV 327&	 D1620601& NO  \\	 
AV 215&	B0 Ia   &   3&	92 &	66\tablenotemark{d}&	AV 327 &	 D1620401& \\	 
AV 216&	B1 III:     &     3&	83 &    &	AV 327&	 D1620501&	 \\
AV 232&	O7 Iaf+     &     5&	76 &	106\tablenotemark{b} 74\tablenotemark{a}&	AV 327&	 P1030201&	 SK 80\\
AV 235&	B0 Iaw     &     5& 	90&    &	 	AV 327&	 P1030301& SK 82 \\	 
AV 238&	O9.5 III     &     5&	71	 &  &	AV 327&	 P1176601& \\	 
AV 242&	B0.7 Iaw     &     5&	83&	54\tablenotemark{d}&	AV 327&	 P1176901& NO \\	 
AV 255&	O8 V     &     6&	188&     &	 	AV 327&	 E5110501& \\	 
AV 261&	O8.5 I     &     6&	87&   &	 	AV 327&	 E5110601& \\	 
AV 264&	B1 I     &     5&	116 &   &	 	AV 327&	 P1177001	& NO \\ 
AV 266&	B1 I     &     3&	61	&  & 	AV 327	& Z9122301	 & NO\\
AV 267&	O8 V     &     7&	268 &  &	 	AV 327&	 E5110701& \\	 
AV 321&	O9 Ib     &     3&	357  &   &	 	AV 327&	 P1150606& \\	 
AV 327&	O9.7 II-Ibw     &     5&	69 &	71\tablenotemark{b} 80\tablenotemark{c}&	AV 327&	 P1176401& \\
AV 332&	O6.5 I +WN3     &     8&	90&     &	 AV 327&	 P1030401, X1050102&	*,SK 108, NO\\
AV 372&	O9 Iabw     &     5&	98&	135\tablenotemark{a}&	AV 327&	 P1176501& \\	 
AV 374&	B2 Ib     &     9&	73&	47\tablenotemark{d}&	AV 327&	 C1600201& NO\\	 
AV 377&	O6 V     &     1&	51&	 	&AV 327&	 B1340101& \\	 
AV 378&	O9 III     &     3&	64 &   &	 	AV 327&	 P1150707&   \\	 
AV 388&	O4 V     &     1&	147 &	163\tablenotemark{a}, 179\tablenotemark{b}&  	AV 327&	 P1175401& \\	 
AV 423&	O9.5 II(n)     &     5& 	163 &	186\tablenotemark{b}&	AV 327&	 P1176701& \\	 
AV 435&	O4 V     &     6&	96 &   &	 	AV 327	& E5111601	&  \\ 
AV 440&	O7 V     &     1&	62 &   &	AV 327&	 E5110901	 & \\
AV 446&	O6.5 V     &     1&	124 &     &	 	AV 327&	 E5111001& \\	 
AV 456&	O9.5 Ib     &     4&	111&     &	 	AV 327&	 Q1070101-4, 0106 &	SK 143\\
AV 462&	B1.5 Ia     &     9&	57 &  	43\tablenotemark{d}&	AV 327&	 A1180301& NO \\	
AV 469&	O8 II     &     1&	79 &	81\tablenotemark{a}&	AV 327&	 P1176301& \\	
AV 472&	B2 Ia     &     9&	73 &	33\tablenotemark{d}&	AV 327&	 C1600301	& NO \\
AV 476&	O2-3 V + comp     &     10& 	65(54)	 &	& AV 327 &	 C0020301& *, NO \\	
AV 488&	B0.5 Iaw     &     11&	87 &	 	&AV 327	& P1176801-3, P1030501&	SK 159\\
%H53-47&	O4 V + O6.5 V      &     8& 	82&    &	 	AV 327&	 F3210101-4, &	* \\
NGC346-MPG324&	O4 V((f))     &     5&	74,118&	70\tablenotemark{b} 120\tablenotemark{c} 105\tablenotemark{a} 70\tablenotemark{e}&	AV 327&	 P2030502, P1175601&	*, NO \\
NGC346-MPG342&	O5-6 V     &     5&	58&	 &	AV 327&	 P2030401, P1175701	& \\
NGC346-MPG355&	O2 III(f*)     &     5&	112&	112\tablenotemark{b} 120\tablenotemark{c} 110\tablenotemark{e}&	SK -66 100&	P1175201	& \\
NGC346-MPG435&	O4 III(n)(f)     &     5&	93 &    &	 	AV 327&	 P2030201, P1175501& \\	
SK 190&	O8 Iaf     &     13&	302 &    &	 	AV 327&	 E5111401& \\	
SK 191&	B1.5 Ia     &     9&   	118 &	92\tablenotemark{d}&	AV 327&	 D1620201& NO \\	
SK 197&	O8.5 Ve     &     12&	166 &   &	 	AV 327&	 E5111501& \\	
\enddata
\tablecomments{Spectral Classifications from (1) \citet{Gar87}, (2 )\citet{Wal00}, (3) \citet{Smi97},(4) \citet{Eva04}, (5) \citet{Wal02}, (6) \citet{Mas95},(7) \citet{Cra82}, (8) \citet{Mas00}, (9) \citet{Duf06}, (10) \citet{Mas05}, (11) \citet{Wal95}, (12) \citet{Cou95}, and (13) \citet{Mas02}.  $V \sin i$ values from (a) \citet{Mok06}, (b) \citet{Pen04}, (c) \citet{Wal00}, (d) \citet{Duf06}, (e) \citet{Bou03}, (f) \citet{Hil03}.  *Additional notes on individual stars are presented in the Appendix.}
\end{deluxetable}

% Table 3 - V sin i for LMC stars
\begin{deluxetable}{llcclcll}
%\rotate
\label{t3}
\tablewidth{0pt}
\tablecaption{LMC Targets}
\tabletypesize{\scriptsize}
\tablecolumns{8}
\tablehead{ 
\colhead{}  &  \colhead{} &  \colhead{} &  \multispan{2}\hfil {$<V \sin i>$}\hfil  & \colhead{} & \colhead{}& \colhead{} \\ 
\colhead{} &\colhead{Spectral}  &  \colhead{} &    \colhead{This paper}  &
 \colhead{Other}  & \colhead{Template} & \colhead{FUSE} & \colhead{} \\
\colhead{Star} & \colhead{Classification} &  \colhead{Ref.} & \colhead{(km~s$^{-1}$)} & \colhead{(km~s$^{-1}$)} & \colhead{Star} 
& \colhead{data sets} 
 &\colhead{Notes} 
}
\startdata 
SK -65 01	& B0.5 I     &     1&	 74	& &	AV 327&	A0490302& NO \\	
SK -65 05 &	O9 II     &     2 &	153  &	 &	AV 327&	E5112301& \\	
SK -65 21	& O9.7 Iab     &     3&  	57(58)	& &	AV 327	& P1020901-4&	*, NO\\
SK -65 22 &	O6 Iaf+     &     3 &	66 &  &	 	AV 327&	P1031002& \\	
SK -65 44& 	O9 V     &     4 &	132 &   &	 	AV 327 &	P11733301-2	& \\
SK -65 47& 	O4 If     &     5 &	208 &  &	 	SK -66 100&	C0020701&  \\	
SK -65 63 &	O9.7 I     &     6 & 	77 &  &	 	AV 327&	A0490701-2, B0861101, M1142001 & \\	
SK -66 01&	B1.5 Ia     &     6 &	78 &  &	 	AV 327&	C1600601& NO \\	
SK -66 18	& O6 V((f))     &     7 & 	100 & 	 82\tablenotemark{a} &	SK -66 100&	A0490101-2	& \\
SK -66 78 &	B1.5 I     &     1 &	65 &  &	 	AV 327&	B1280301& NO\\	
SK -66 97 &	B1 Iab     &     8 & 	159 &   &	 	SK -66 100 &	E9570101&	Hen S35, NO \\
SK -66 100 &	O6 II(f)     &     3 &	67 &	 84\tablenotemark{a}&	AV 327&	P1172303	& \\
SK -66 106&	B2 Ia     &     6 &	78 &   &	 	AV 327&	B1280901& NO \\	
SK -66 118&	B2 Ia     &     6 &	167 &  &	 	AV 327&	C1600701& NO \\	
SK -66 169 &	O9.7 Ia+     &     3 &	71&	72\tablenotemark{b} &	AV 327&	P1173801&	HD 269889\\
SK -66 171 &	O9 Ia     &     6 &	90 &  &	 	AV 327&	E5114102& \\	
SK -66 172 &	O2 III(f*)+OB     &     3 &	101 &  &	 	SK -66 100&	P1172201&	*, NO\\
SK -66 185 &	B0 Iab     &     2 &	53 &  &	 	AV 327&	A0490902& \\	
SK -67 05	 & O9.7 Ib     &     3 &	87 &  &	 	AV 327&	P1030703-4	& \\
SK -67 14 &	B1.5 Ia     &     3 &	88 &  &	 	AV 327& 	P1174201-3	& NO \\
SK -67 18	& O3 + O8-B0 I     &     9 &	80(67)	& &	SK -66 100&	D0980801&	*, NO\\
%SK -67 22	& O2 If*     &     10 &    &   &	 	 	AV 327&	C0020401& \\	
SK -67 28& 	B0.7 Ia     &     6 &	76 &  &	 	AV 327&	A0490201-2& NO \\	
SK -67 38& 	O8.5 V     &     7 &	100 &   &	 	AV 327& 	E5112601& \\	
SK -67 46& 	B1.5 I     &     1 &	71& &	 	AV 327&	A0491501& NO \\	
SK -67 69& 	O4 III(f)     &     3	&121 &   &	 	SK -66 100 &	P1171701, 1703& \\	
SK -67 76	& B0 Ia     &     11 &	80 &  &	 	AV 327&	P1031201& \\	
SK -67 100 &	B1 Ia     &     6 &	88 &  &	 	AV 327&	B1280401&	HD 269504, NO\\
SK -67 101 &	O8 II((f))     &     3 &	103 &	101\tablenotemark{b} &	AV 327&	P1173401-3	& \\
%SK -67 105 &	O4f + O6 V 	186.7	 	AV 327	D1530101Ñ6	
SK -67 106&	B0 I     &     12 &	99 & 	135\tablenotemark{b}&	AV 327&	A1110101& \\	
SK -67 107&	B0 I     &     13&	82 &	103\tablenotemark{b}&	AV 327 &	A1110201& \\	
SK -67 111&	O6 Ia(n)fp var     &     3&	259&	209\tablenotemark{b}&	AV 327&	C1550208-9, C1550214-5	& \\
SK -67 118&	O7 V     &     7&	61 &	 	& AV 327& 	E5113401& \\	
SK -67 119&	O7 III(f)     &     7 &	235 &	 	&AV 327&	E5113501	& \\
SK -67 150&	B1 Ia     &     4&	66 &  & 	AV 327&	B1281001& NO \\	
SK -67 166&	O4 If+     &     3&	97 &	157\tablenotemark{b}, 97\tablenotemark{a}&	SK -66 100&	A1330101-162, A1330180&	*\\
SK -67 167&	O4 Inf+     &     3&	155 &  &	 	SK -66 100&	P1171901-2& \\	
SK -67 169&	B1 Ia     &     3&	62 &  &	 	AV 327&	P1031603& NO \\	
SK -67 174&	O8 V     &     7&	72 &  &	 	AV 327&	D0980101	&  \\
SK -67 176&	O7 Ib(f)     &     6&	54	&  & 	AV 327	&D0980201	&  \\
SK -67 181&	B0.5 I     &     11& 	99 &  &	 	AV 327	&B0900101	& NO \\
SK -67 191&	O8 V     &     2&	100 &  &	 	AV 327&	P1173101-2	& \\
%SK -67 211&	O2 III(f*)     &     3&	 	&173\tablenotemark{b}&	AV 327&	P1171601, 1603&	HD \AA\\
SK -67 256&	B1 Ia     &     6&	66&  &	 	AV 327&	B1280201& NO \\	
SK -67 266&	Ofpe/WN9     &     14&	113 &	144\tablenotemark{b}&	SK -66 100&	B0900201& \\	
SK -68 03&	O9 I     &     2&	70 &  &	 	AV 327&	A0490401-2	& \\
SK -68 16	& O7 III     &     2& 	100 &	 &	AV 327&	E5112201& \\	
SK -68 41&	B0.5 Ia     &     6&	77 &   &	AV 327& 	P1174101-2	& NO \\
SK -68 52 &	B0 Ia     &     3&	70(55)&   &	 	AV 327&	P1174001& 	 *, NO\\
SK -68 75 &	B1 I     &     1&	78 &  &	AV 327&	A0490501-2	& NO \\
SK -68 111 &	BN1 Ia     &     15&	72 &  &	 	AV 327&	B1280501&	HD 269668, NO\\
SK -68 129&	B1 I     &     5 &	96 &	 	& AV 327&	B0860501& NO \\	
SK -68 135&	ON9.7 Ia+     &     16 &	68 &70\tablenotemark{b}&	AV 327&	P1173901&	HD 269896\\
SK -68 137&	O3 III(f*)     &     16&	107 &  &	 	SK -66 100&	E5114201& \\	
SK -68 155&	B0.5      &     11& 	96 &   & 	SK -66 100&	B0860701& NO \\	
SK -68 171&	B1 Ia     &     6 &	71 &  & 	AV 327	&A0490801	& NO \\
SK -69 43 &	B0.5 Ia     &     6 &	102 &  &	 	AV 327&	B1280601&	HD 268809, NO\\
SK -69 50	& O7 If     &     7 &	178 &  &	 	AV 327&	E5112001& \\	
SK -69 52	& B2 Ia     &     6 &	105	 & &	AV 327	& P1174301	& NO \\
SK -69 59 &	B0 Ia     &     11 &	111 & & 	 	AV 327& 	P1031103	& \\
SK -69 94 &	A0e/LBV &	& 127	&   & 	AV 327&	Z9050201	 & *,S Dor = HD 35343, NO \\
SK -69 104 &	O6 Ib(f)     &     3&	117 &	97\tablenotemark{b} &	AV 327&	P1172401 & HD 269357 \\	
SK -69 124&	O9 Ib     &     3& 	96 &  &	 	AV 327&	P1173601-2	& \\
SK -69 223 &	WC4(+OB)     &     17& 	84 &  &	 	AV 327	& P2150201-2	& *HD 38029, Brey 67, NO\\
SK -69 237&	B1 Ia     &     15 & 	86	&  &	AV 327&	B1280701& NO	\\
SK -69 279&	O9 f     &     2&	84 & &	AV 327&	B0860801& \\	
SK -70 13	& 	O9 V     &     7&	84 & &	 	AV 327	& D0981001& \\	
SK -70 32 &	O9.5 II     &     18 &	97 &  &	 	AV 327&	E5112501& \\	
SK -70 60	&	O4-5 V:n     &     13 &	256 &  &	 	SK -66 100&	P1172001& \\	
SK -70 69	&	O5 V     &     3 &	98 &	 131\tablenotemark{a} &	AV 327&	P1172101& \\	
SK -70 78 &	B0.7 Ia     &     6 &	107 & &	 	AV 327&	B1280801	&HD 269074\\
SK -70 79	&	B0 III     &     7 &	86 &  & 	AV 327& 	B0900301& \\	
SK -70 85 &	B0 I     &     1 &	63 &  &	 	AV 327&	A0491301, A0491303& \\	
SK -70 91	& 	O2 III(f*)+OB     &     3 &	99(78)&  &	 	SK -66 100&	P1172501&	*, NO\\
SK -70 115&	O7 Ib     &     4 &	164 & &	 	AV 327&	P1172601& \\	
SK -70 120&	B1 Ia     &     6 &	75 & &	 	AV 327	& A0491001-2	& NO \\
SK -71 08	&	 O9 II     &     2 &	75 & &	 	AV 327&	A0491401-2	& \\
SK -71 19	& 	O7 V     &     4 &	242  &  &	 	AV 327&	E5113001	& \\
SK -71 41	& 	O8.5 I     &     7 &		53 &  &	 	AV 327	& D0980501	& \\
SK -71 45 &	O4-5 III(f)     &     3 &	117 &	125\tablenotemark{b} &	SK -66 100&	P1031501-4	&HD 269676\\
SK -71 46	& 	O4 If     &     7 &	79 & & 	SK -66 100&	D0980601& \\	
SK -71 50	&	O6.5 III     &     2 &	296 &  &	 	AV 327&	A0491201	& \\
BI 12	&	O6 III     &     4 &	116 & & 	AV 327& 	E5111802	& \\
BI 13	&	O6.5 V     &     2&	 118 &  &		SK -66 100&	E5111901& \\	
BI 130 &	O8.5 V((f))     &     7 &	95 &  &	 	AV 327& 	E5112902& \\	
BI 170 &	O9.5 Ib     &     3 &	84&  &	 	AV 327	& P1173701	& \\
BI 173&	 O8 II:     &     3 &	116 & &	 	SK -66 100&	P1173201-2	& *, NO\\
BI 208&	O7 V     &     2 &	196 &  &	 	AV 327	& P1172702-5& \\	
BI 237& 	O2 V((f*))     &     10 &	83 &	 126\tablenotemark{a} &	SK -66 100&	E5114001& \\	
BI 253&	O2 V((f*))     &     10 &	155 &	 191\tablenotemark{a} &	SK -66 100&	C0020601-2	& \\
BI 272&	O7 III      &     19 &	212 &	& 	AV 327&	P1172901-2	& \\
HV 2241&		O7 III + B0 III     &     20 &	71 & & 	AV 327&	F9270603& 	*, NO\\
%HV 2274	B1-2 IV-III + B1-2 IV-III 	192.7	 	AV 327	B0770201	
LH47-338&	O7 V((f))     &     5&	99 &	 &	AV 327&	P2320101	&N44C Ð Star 2\\
LMC X-4 &	O8 III     &     21 &	322 &  &	 	AV 327&	A0440102&	*, NO \\
LMC2-703 &	O9.5 V     &     7 &	95 &   &	 	AV 327&	D0981401 & D301-1005 \\	
LMC2-755 &	O8 V     &     7 &	57 &   &	 	AV 327	& D0981501 & D301-NW8 \\	
PGMW 1486 &	O6.5 V     &     22 &	93 &  &	 	SK -66 100&	D0300101& \\	
PGMW 3053&	O5.5 I-III(f)     &     22 &	110 &  &	 	SK -66 100&	B0100101& \\	
PGMW 3070&	O6 V     &     22 &	72 &  &	 	AV 327&	B0100201& \\	
PGMW 3073&	O6.5 V      &     22 &	71 &  &	AV 327&	D0300201& \\	
PGMW 3102&	O7 V     &     22& 	68 &  	& AV 327&	D0300302& \\	
PGMW 3120&	O5.5 V((f*))     &     22 & 	111 & 95\tablenotemark{b}	 &	SK -66 100&	B0100301	& \\
PGMW 3157&	BC1 Ia     &     22 &	88 & & 	AV 327&	B0100401& NO \\	
PGMW 3168&	O7 II(f)     &     22 &	78 &	 96\tablenotemark{a} &	SK -66 100&	B0100501& \\	
PGMW 3204&	O6-7 V      &     22 &	78 &	 130\tablenotemark{a} &	SK -66 100&	D0300402	& \\
PGMW 3209&	O3 III(f*)+OB     &     22 &	166 &	 	& AV 327& 	B0100601& *, NO\\	
PGMW 3223&	O8.5 IV     &     22 &	147&  & 	SK -66 100 & 	B0100701& \\	
PGMW 3264&	O3-6 V     &     22 &	56 & &	SK -66 100	& B0100801	& \\
\enddata
\tablecomments{Spectral Classifications from (1) \citet{Jax01}, (2) \citet{Con86}, (3) \citet{Wal02}, (4) \citet{Smi99},  (5) \citet{Mas00}, (6) \citet{Fit88}, (7) \citet{Mas95},  (8) \citet{Gum95}, (9) \citet{Koe03}, (10) \citet{Mas05}, 
(11) \citet{Rou78}, (12) \citet{Feh82}, (13) \citet{Rob03}, (14) \citet{Boh89}, (15) \citet{Fit91}, (16) \citet{Wal77}, (17) \citet{Bar01}, (18) \citet{Mas02}, (19) \citet{Pla98}, (20) \citet{Ost01}, (21) \citet{Neg02}, and (22) \citet{Par92}.  $V \sin i$ values from (a) \citet{Mok07} and (b)\citet{Pen04}.   *Additional notes on individual stars are presented in the Appendix.}
\end{deluxetable}

\placetable{t4}      % Table 4 - V sin i for SMC stars from other sources
\begin{deluxetable}{lccr}
%\rotate
\label{t4}
\tablewidth{0pc}
\tablecaption{SMC Stars with Previous $V \sin i$ Values}
\tabletypesize{\scriptsize}
\tablecolumns{4}
\tablehead{ 
\colhead{}  &  \colhead{Spectral} & \colhead{$<V \sin i>$}  & \colhead{}  \\ 
\colhead{Star} &\colhead{Classification}  &    \colhead{(km~s$^{-1}$)}  &
 \colhead{Reference} }
\startdata 
346-010&O7 IIIn((f))&313&M06\\
346-012&B1 Ib&29&M06\\
346-018&O9.5 IIIe&138&M06\\
346-022&O9 V&55&M06\\
346-025&O9 V&138&M06\\
346-031&O8 Vz&18&M06\\
346-033&O8 V&188&M06\\
346-046&O7 Vn&340&M06\\
346-050&O8 Vn&357&M06\\
346-051&O7 Vz&18&M06\\
346-066&O9.5 V&129&M06\\
346-077&O9 V&177&M06\\
346-090&O9.5 V&188&M06\\
346-093&B0 V&187&M06\\
346-097&O9 V&22&M06\\
346-107&O9.5 V&55&M06\\
346-112&O9.5 V&143&M06\\
330-013&O8.5 III((f))&73&M06\\
330-052&O8.5 Vn&291&M06\\
NGC346-MPG368&O4-5 V((f))&55&P04\\
NGC346-MPG113&OC6 Vz&$<$40&P04\\
AV 243&O6 V&62&P04\\
NGC346-MPG487&O6.5 V&65&P04\\
AV 220&O6.5f?p&$<$40&P04\\
NGC346-MPG682&O8 V&71&P04\\
NGC346-MPG12&O9.5-B0 V&67&P04\\
AV 170&O9.7 III&54&P04\\
\enddata
\tablecomments{M06 = \citet{Mok06} and P04 = \citet{Pen04}}
\end{deluxetable}

\placetable{t5}      % Table 5 - V sin i for LMC stars from other sources
\begin{deluxetable}{lccr}
%\rotate
\label{t5}
\tablewidth{0pc}
\tablecaption{LMC Stars with Previous $V \sin i$ Values}
\tabletypesize{\scriptsize}
\tablecolumns{4}
\tablehead{ 
\colhead{}  &  \colhead{Spectral} & \colhead{$<V \sin i>$}  & \colhead{}  \\ 
\colhead{Star} &\colhead{Classification}  &    \colhead{(km~s$^{-1}$)}  &
 \colhead{Reference} }
\startdata 
N11-004&OC9.7 Ib&81&M07\\
N11-008&B0.7 Ia&83&M07\\
N11-026&O2 III(f*)&109&M07\\
N11-029&O9.7 Ib&77&M07\\
N11-031&ON2 III(f*)&116&M07\\
N11-033&B0 IIIn&256&M07\\
N11-036&B0.5 Ib&54&M07\\
N11-038&O5 II(f+)&145&M07\\
N11-042&B0 III&42&M07\\
N11-045&O9 III&105&M07\\
N11-051&O5 Vn((f))&333&M07\\
N11-058&O5.5 V((f))&85&M07\\
N11-060&O3 V((f*))&106&M07\\
N11-061&O9 V&87&M07\\
N11-065&O6.5 V((f))&83&M07\\
N11-066&O7 V((f))&71&M07\\
N11-068&O7 V((f))&54&M07\\
N11-072&B0.2 III&14&M07\\
N11-087&O9.5 Vn&276&M07\\
N11-123&O9.5 V&110&M07\\
HD 269810&O2 III(f*)&173&P04\\
SK -69 212&O5 III(f)&210&P04\\
SK -67 51&O6.5 III&105&P04\\
CD -68 264&O8 V&139&P04\\
LH58-52A&O8 II&110&P04\\
LH58-19A&O9 II&178&P04\\
\enddata
\tablecomments{M07 = \citet{Mok07} and P04 = \citet{Pen04}}
\end{deluxetable}

\placetable{t6}      % Table 6 - K-S statistics
\begin{deluxetable}{lcclcccr}
%\rotate
\label{t6}
\tablewidth{0pc}
\tablecaption{Kolmogorov-Smirnov Statistics}
\tabletypesize{\scriptsize}
\tablecolumns{8}
\tablehead{ 
\colhead{} & \colhead{$n_1$} & \colhead{$<V \sin i>$} & \colhead{} & \colhead{$n_2$} & \colhead{$<V \sin i>$} &\colhead{} & \colhead{$p$}\\
\colhead{Sample 1}  &  \colhead{stars} & \colhead{(km~s$^{-1}$)} & \colhead{Sample 2}  & \colhead{stars} & \colhead{(km~s$^{-1}$)} &\colhead{$D$} & \colhead{(\%)} }
\startdata 
\vspace{-1pt} \\

\multispan{8}\hfil Unevolved Vs. Evolved Within Each Environment \hfil \\
\vspace{-8pt} \\

\multispan{8}\hrulefill \\
\vspace{-1pt} \\

Galactic unevolved & 79 & 129.8 & Galactic evolved & 56 &109.5 & 0.296 & 0.5 \\
LMC unevolved & 36 & 132.0 & LMC evolved & 42 & 118.6 & 0.151 & 74 \\
SMC unevolved & 32 & 116.2 & SMC evolved & 19 &101.8 & 0.289 & 23 \\

\vspace{-1pt} \\

\multispan{8}\hfil Unevolved Samples \hfil \\

\multispan{8}\hrulefill \\
\vspace{-1pt} \\

LMC unevolved & 36 & 132.0 & Galactic unevolved & 79 & 129.8 & 0.180 & 37 \\
SMC unevolved & 32 & 116.2  & Galactic unevolved & 79 & 129.8 &0.172&  47 \\
SMC unevolved & 32 & 116.2 & LMC unevolved & 36 & 132.0 & 0.295 & 8 \\
\vspace{-1pt} \\

\multispan{8}\hfil Evolved Samples \hfil \\
\multispan{8}\hrulefill \\
\vspace{-1pt} \\

LMC evolved & 42 & 118.6  & Galactic evolved & 56 & 109.5 &  0.321 & 1 \\
SMC evolved & 19 & 101.8 & Galactic evolved & 56 & 109.5 & 0.297 &  13 \\
SMC evolved & 19 & 101.8 & LMC evolved & 42 & 118.6 & 0.137  & 95 \\

\enddata
\end{deluxetable}

%%%%%%%%%%%%%%%%%%%%%%%%%%%%%%%%%%%%%%%%%%%%%%%%%%%%%%%%%%%%%%%%%%%%%%%%%%%
% Figures

\placefigure{f1}      % Figure 1
\setcounter{figure}{0}
\begin{figure}
\epsscale{1.0}
\plotone{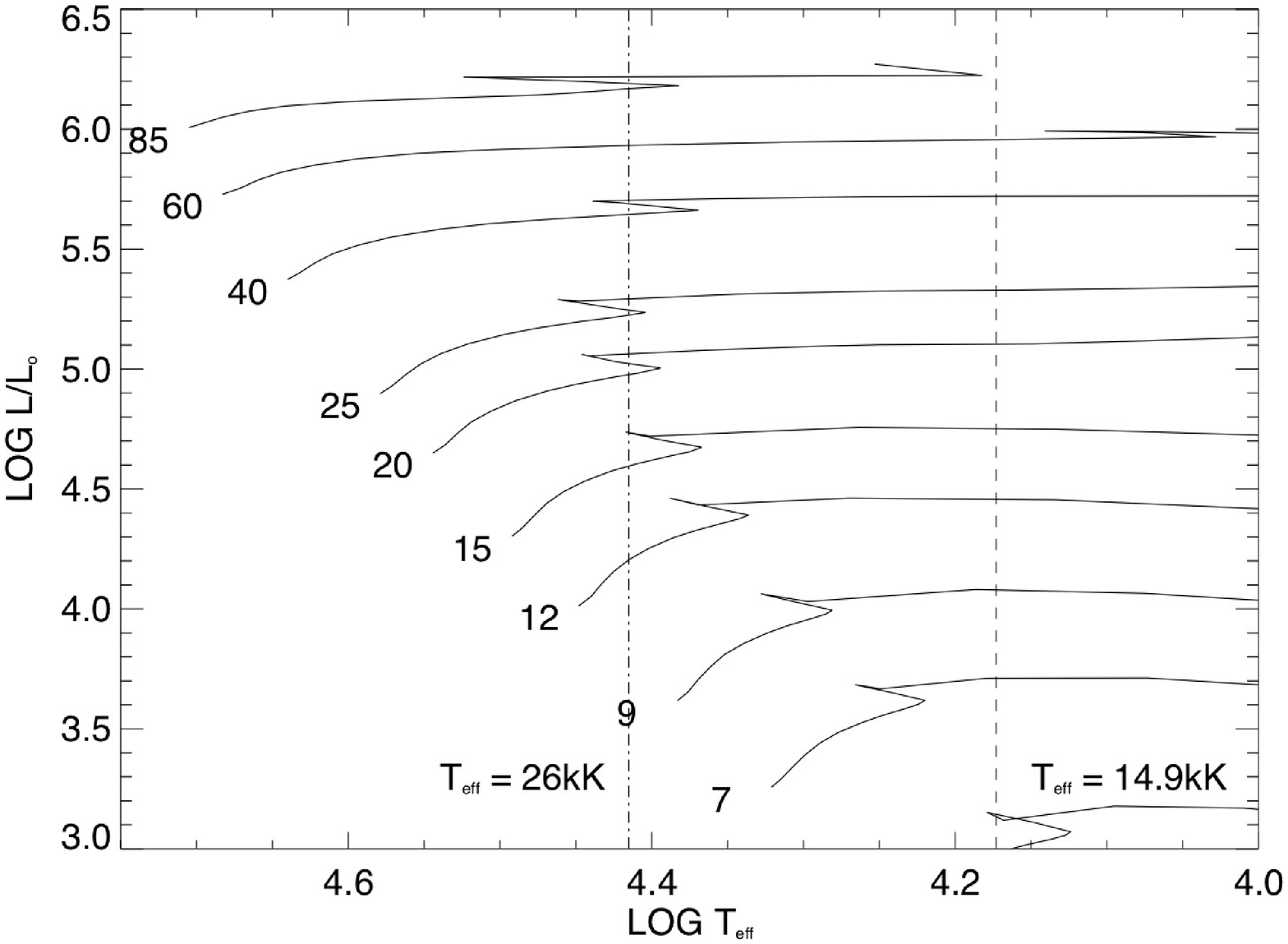}
\caption{ Evolutionary tracks from \citet{Sch92} with an overlain lines at $T_{\rm eff} = 26000$K and $T_{\rm eff} = 14900$K.  The first corresponds to spectral class O9.7 I by \citet{Cro02}, presumably the lower limit of effective temperatures to be assigned to O spectral type stars.  All stars to the right of this line will be typed as B types or cooler.  The second overlaid line corresponds to the temperature for spectral classification B3 I \citep{Boh81}.  Stars with this temperature may have progenitors with initial masses as low as $7 M_\odot$.}
\label{f1}
\end{figure}

\placefigure{f2}      % Figures 2a - 2d
\begin{figure}
\epsscale{1.0}
\plottwo{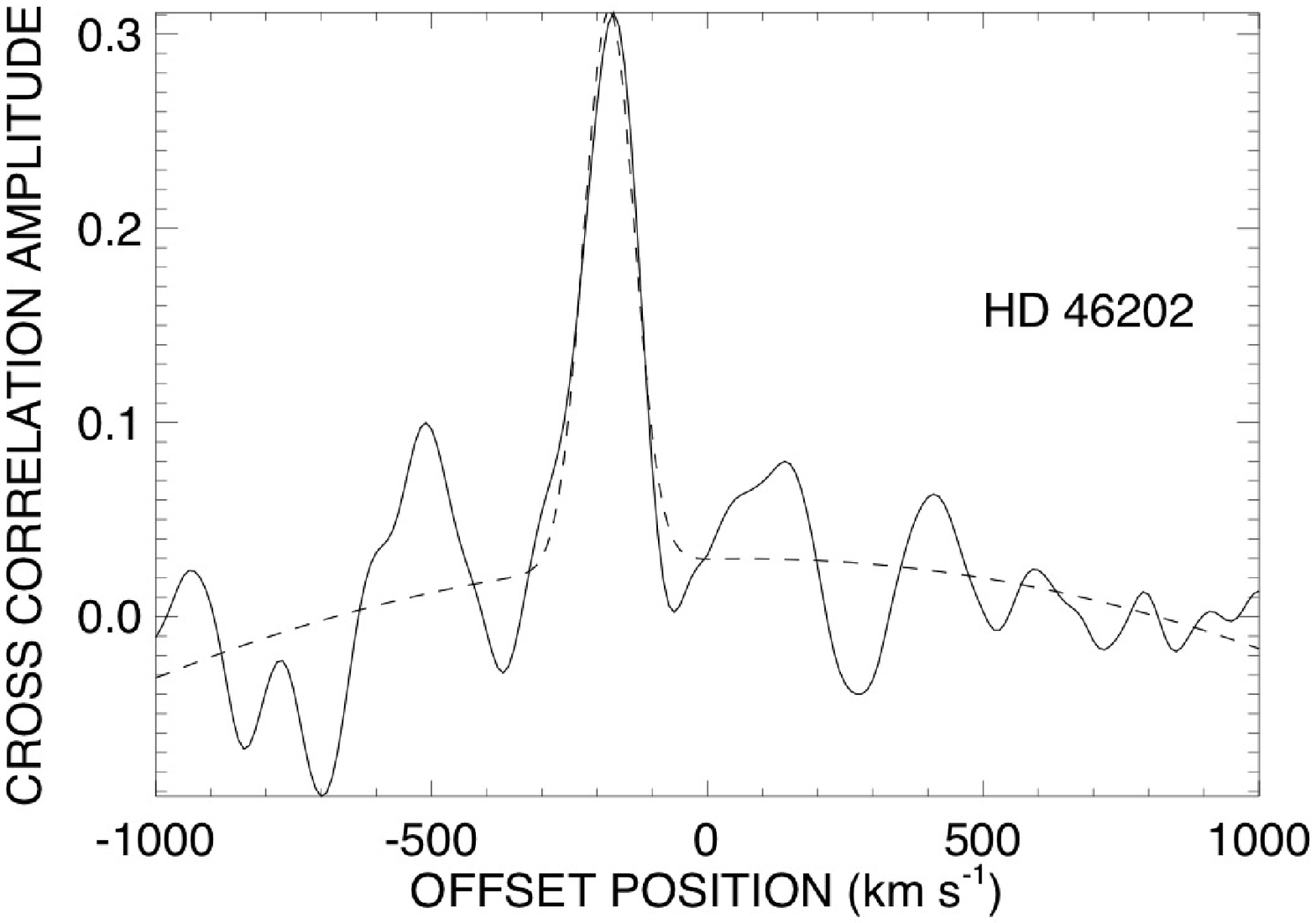}{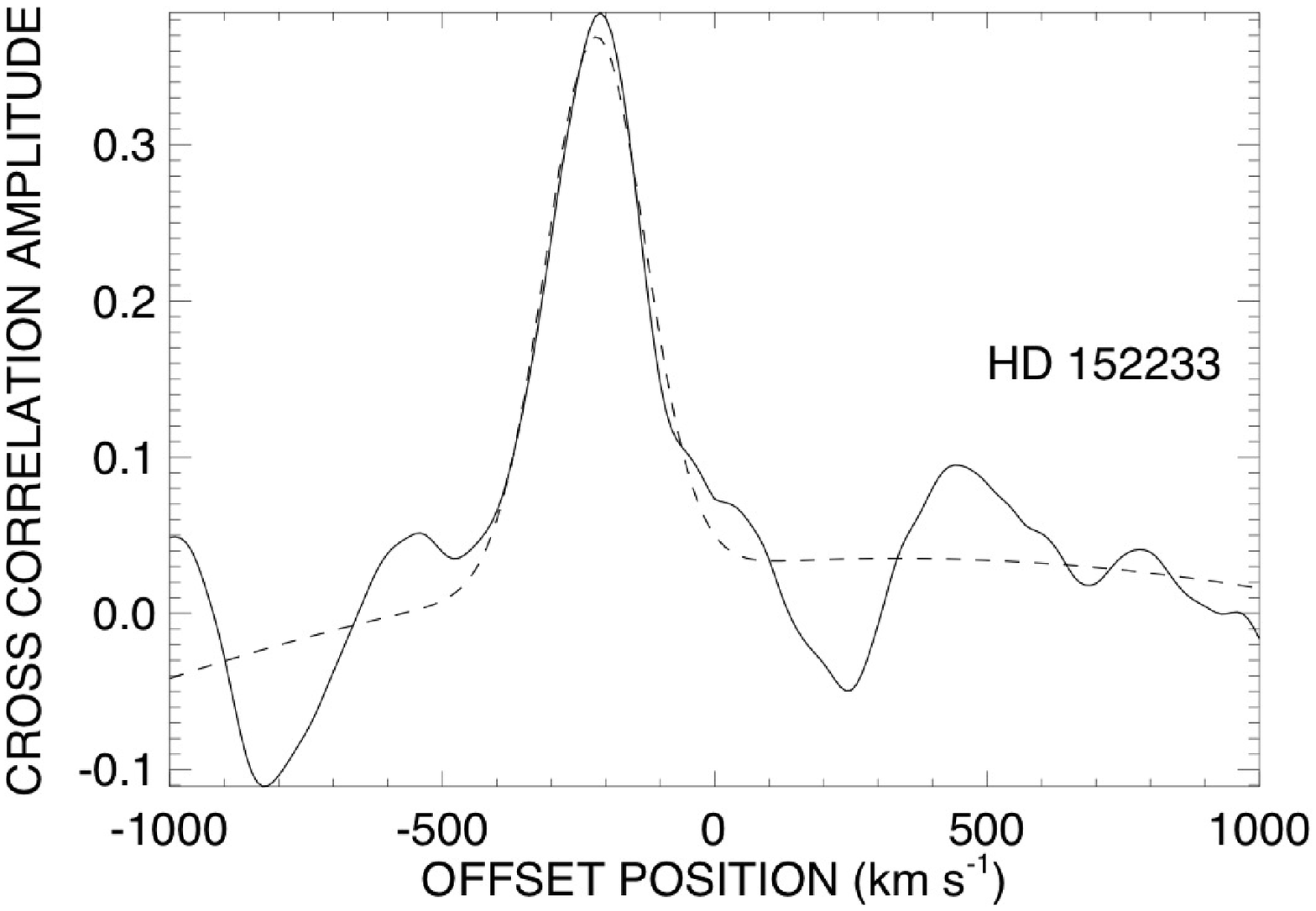}
\plottwo{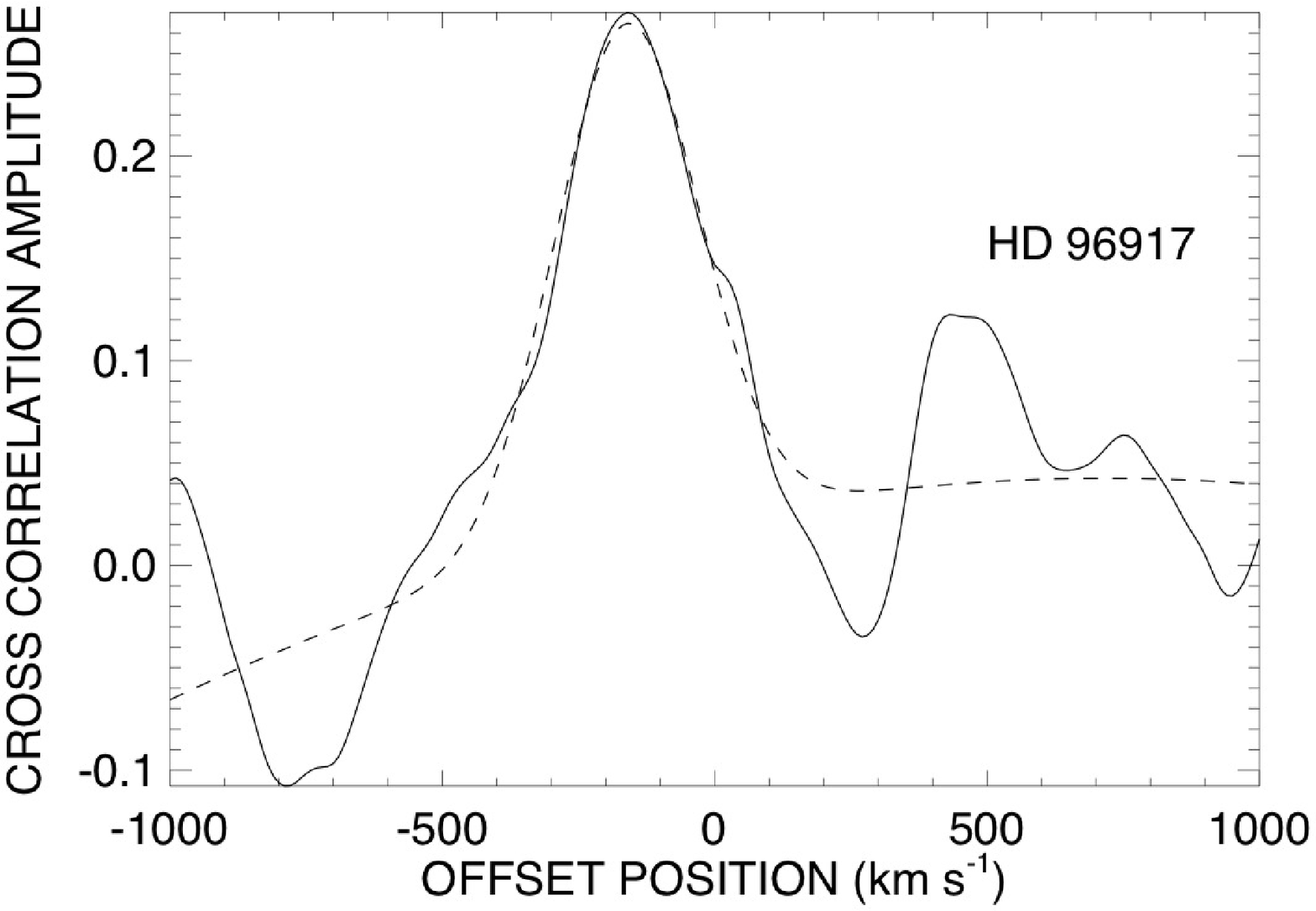}{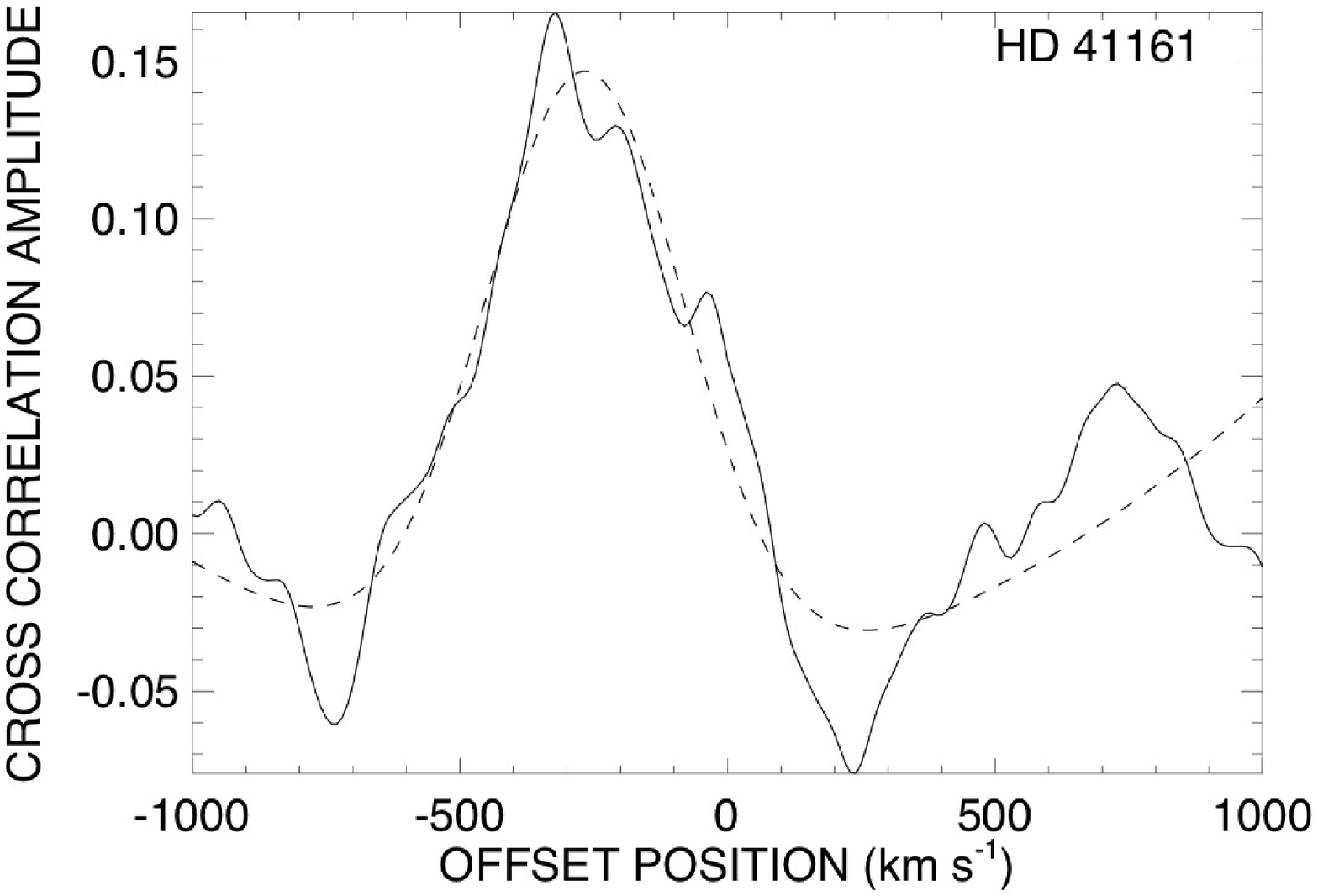}
\caption{The cross-correlation functions and Gaussian fits for the Galactic stars 
(a) HD~46202 (O9 V; $V \sin i = 37$ km~s$^{-1}$), 
b) HD~152233 (O6 III:(f)p; $V \sin i = 104$ km~s$^{-1}$), 
(c) HD~96917 (O8.5 Ib(f); $V \sin i = 176$ km~s$^{-1}$), (d) HD~41161 (O8 Vn; 
$V \sin i = 304$ km~s$^{-1}$).  The bumpy nature of the ccf for HD~41161 may be 
indicative of non-radial pulsations.}
\label{f2}
\end{figure}

\placefigure{f3}      % Figures 3a - 3b
\begin{figure}
\epsscale{1.0}
\plottwo{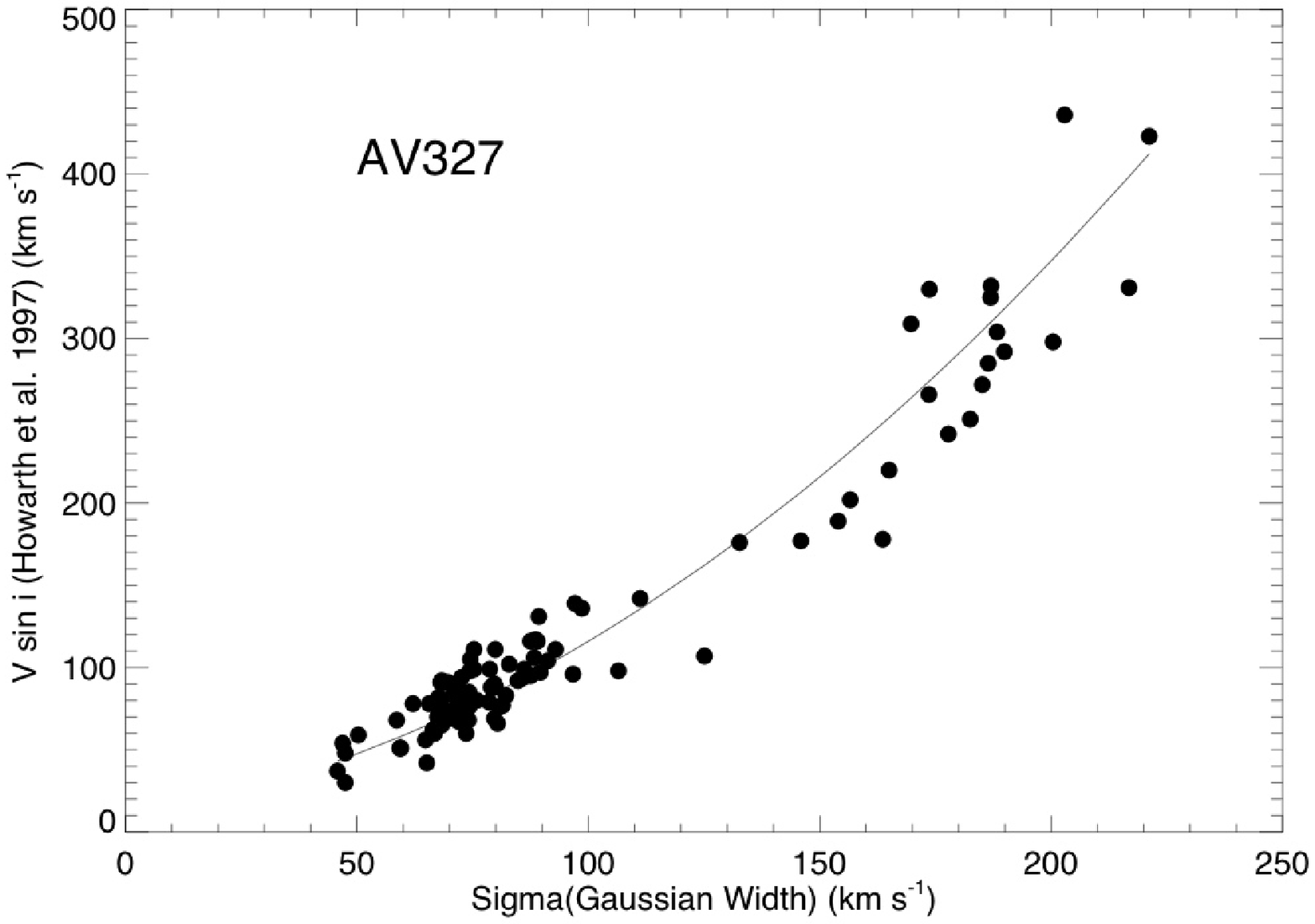}{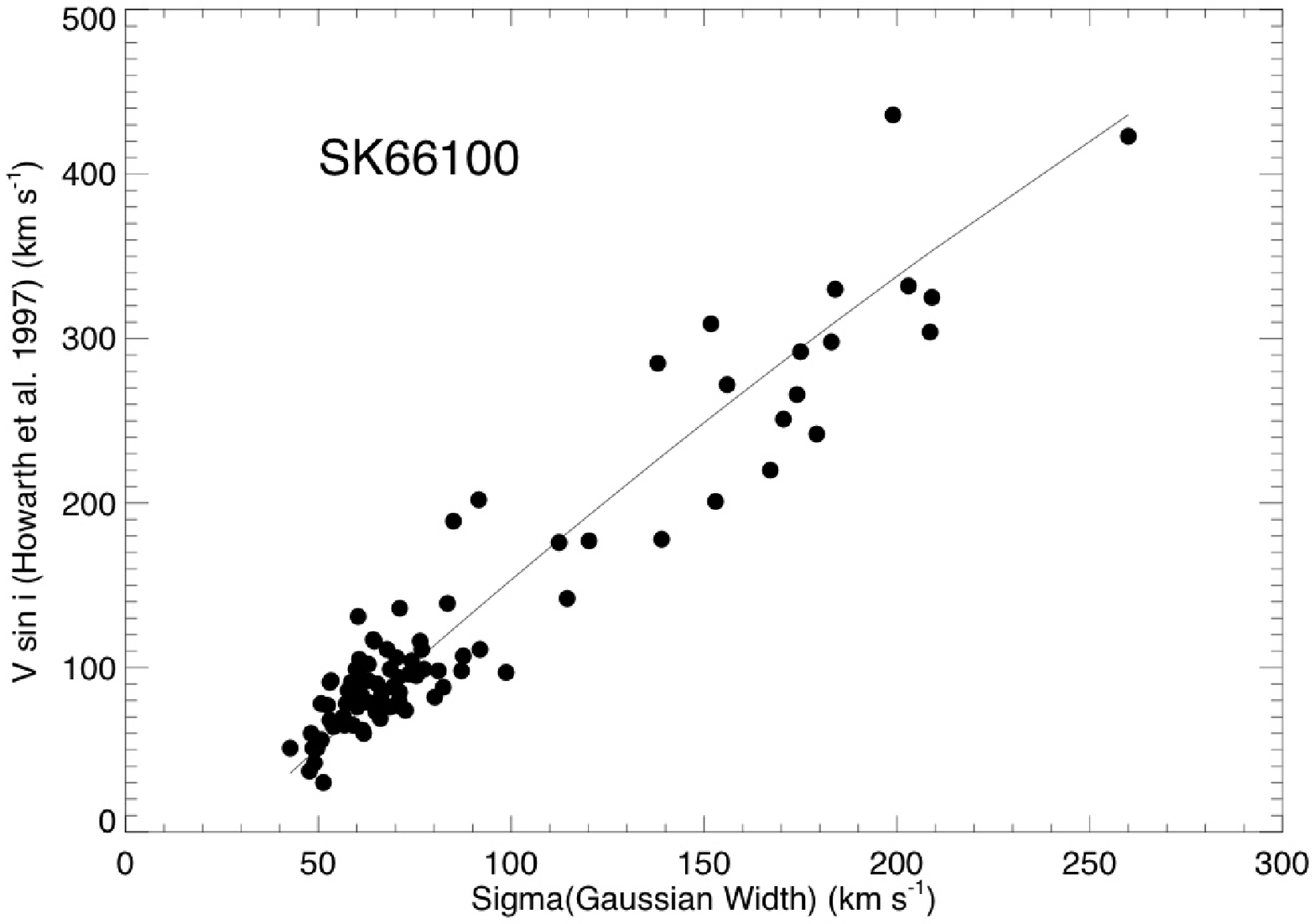}
%\plotone{AV440_wds.pdf}
\caption{The projected rotational velocities of \citet{Pen96} or \citet{How97} from IUE spectra vs. the Gaussian width of the CCF's from the template (a) AV 327 (O9.5 II-Ibw) and (b) SK -66 100 (O6 II(f)). }
\label{f3}
\end{figure}

\placefigure{f4}      % Figures 4a - 4b
\begin{figure}
\epsscale{1.0}
\plottwo{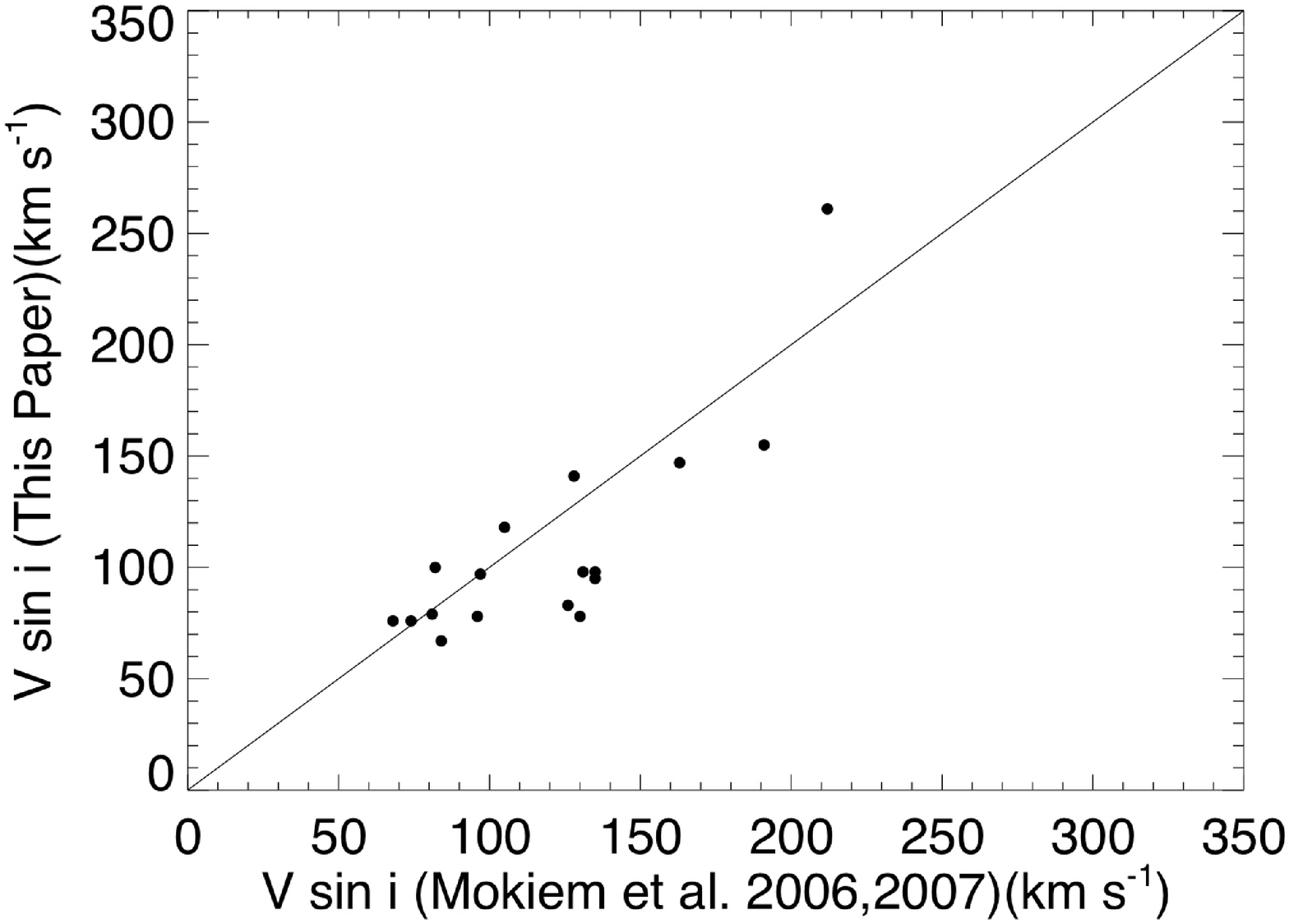}{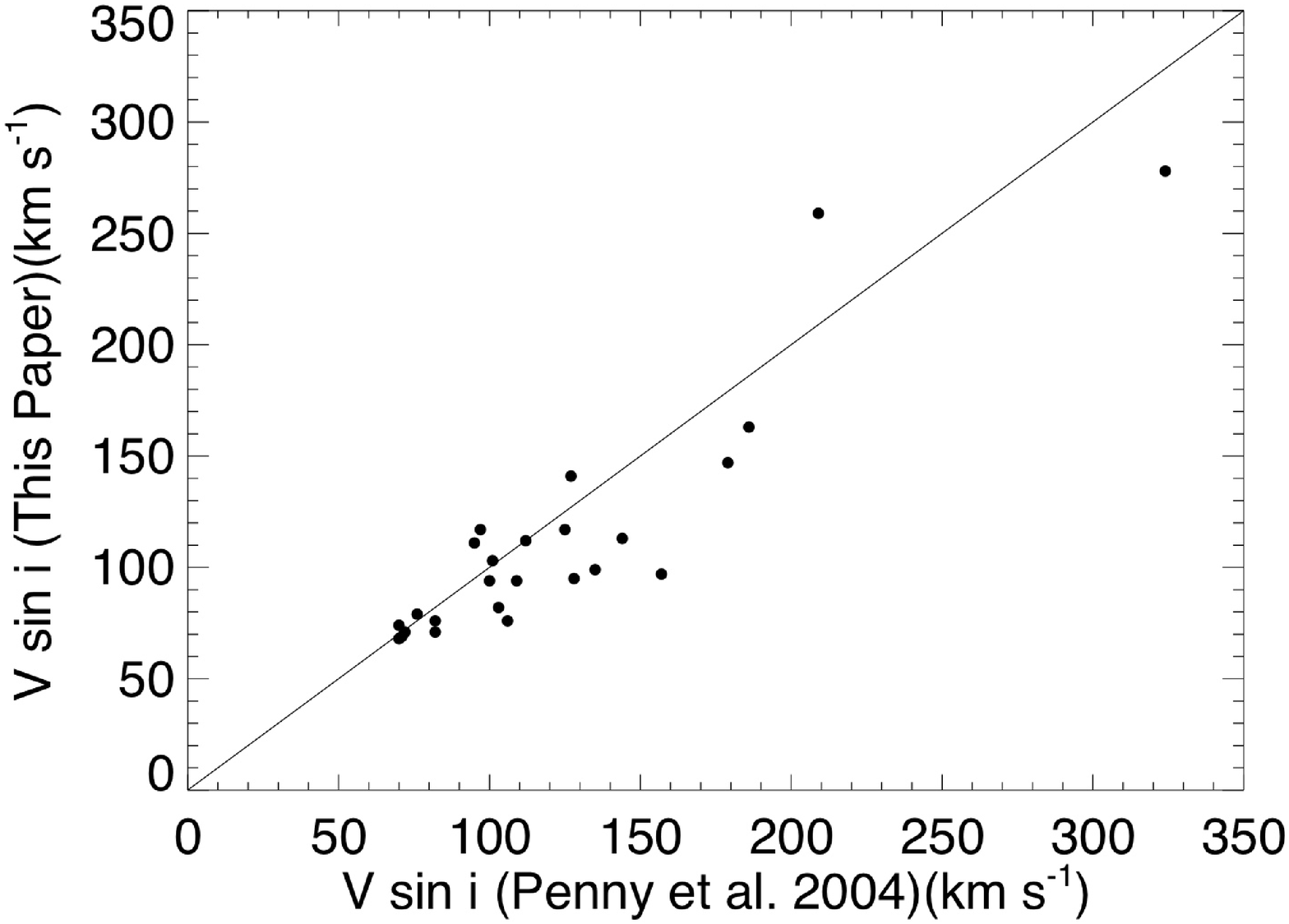}
%\plotone{AV440_wds.pdf}
\caption{Projected rotational velocities from Tables 2 and 3 versus (a) \citet{Mok06,Mok07} and (b) \citet{Pen04} for those targets in common.}
\label{f4}
\end{figure}

\placefigure{f5}      % Figure 5
\begin{figure}
\epsscale{1.0}
\plotone{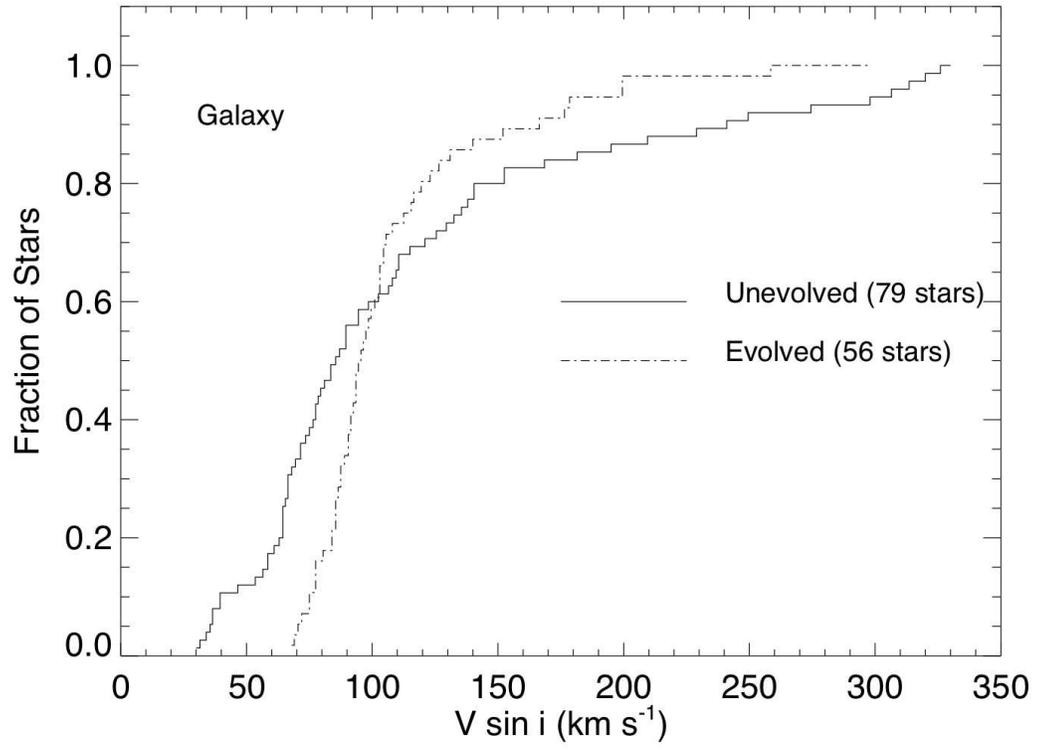}
\caption{Cumulative distribution functions of $V \sin i$ values for unevolved and evolved stars in the Milky Way.}
\label{f5}
\end{figure}

\placefigure{f6}      % Figure 6
\begin{figure}
\epsscale{1.0}
\plotone{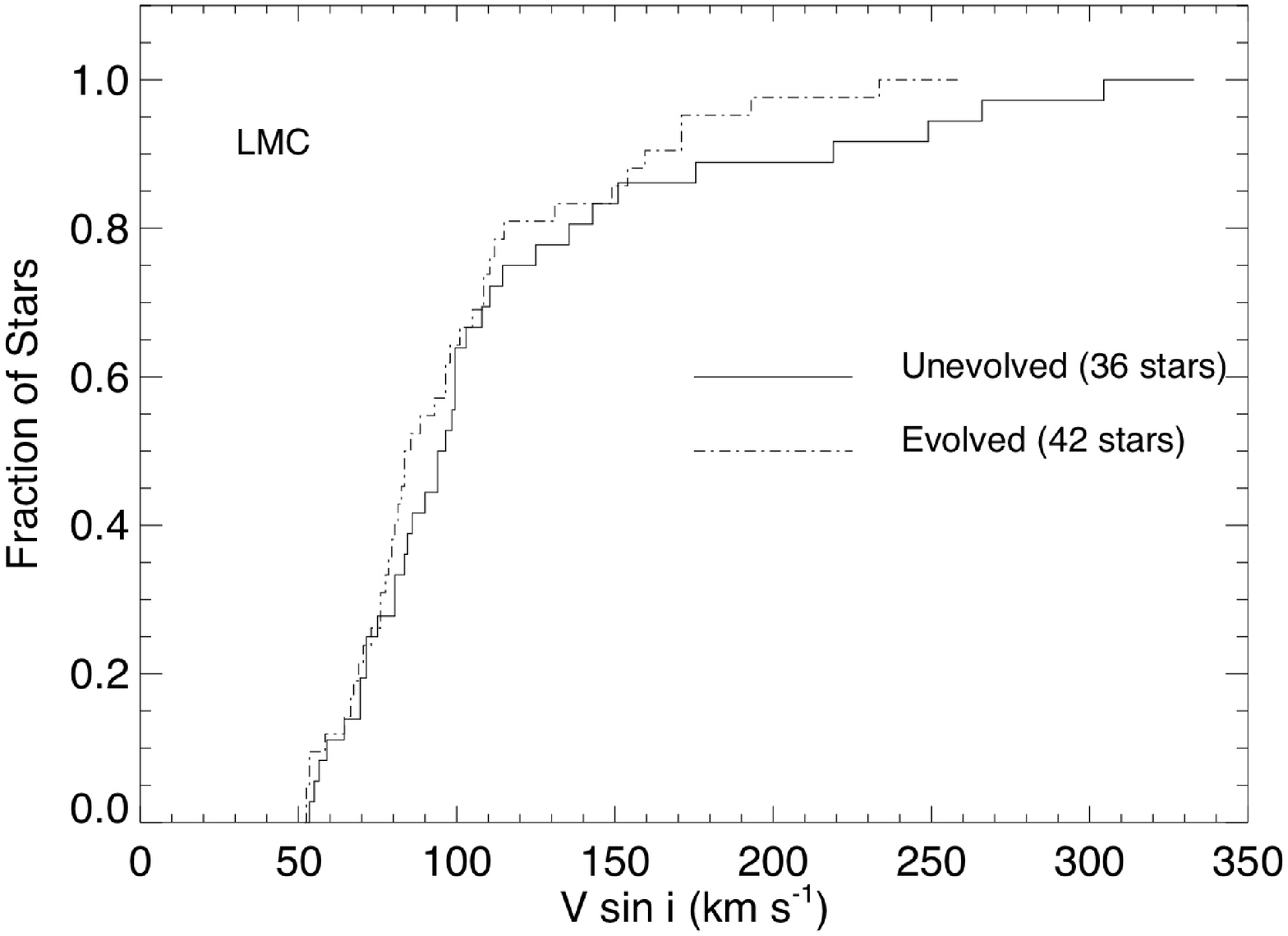}
\caption{Cumulative distribution functions of $V \sin i$ values for unevolved and evolved stars in the LMC.}
\label{f6}
\end{figure}

\placefigure{f7}      % Figure 7
\begin{figure}
\epsscale{1.0}
\plotone{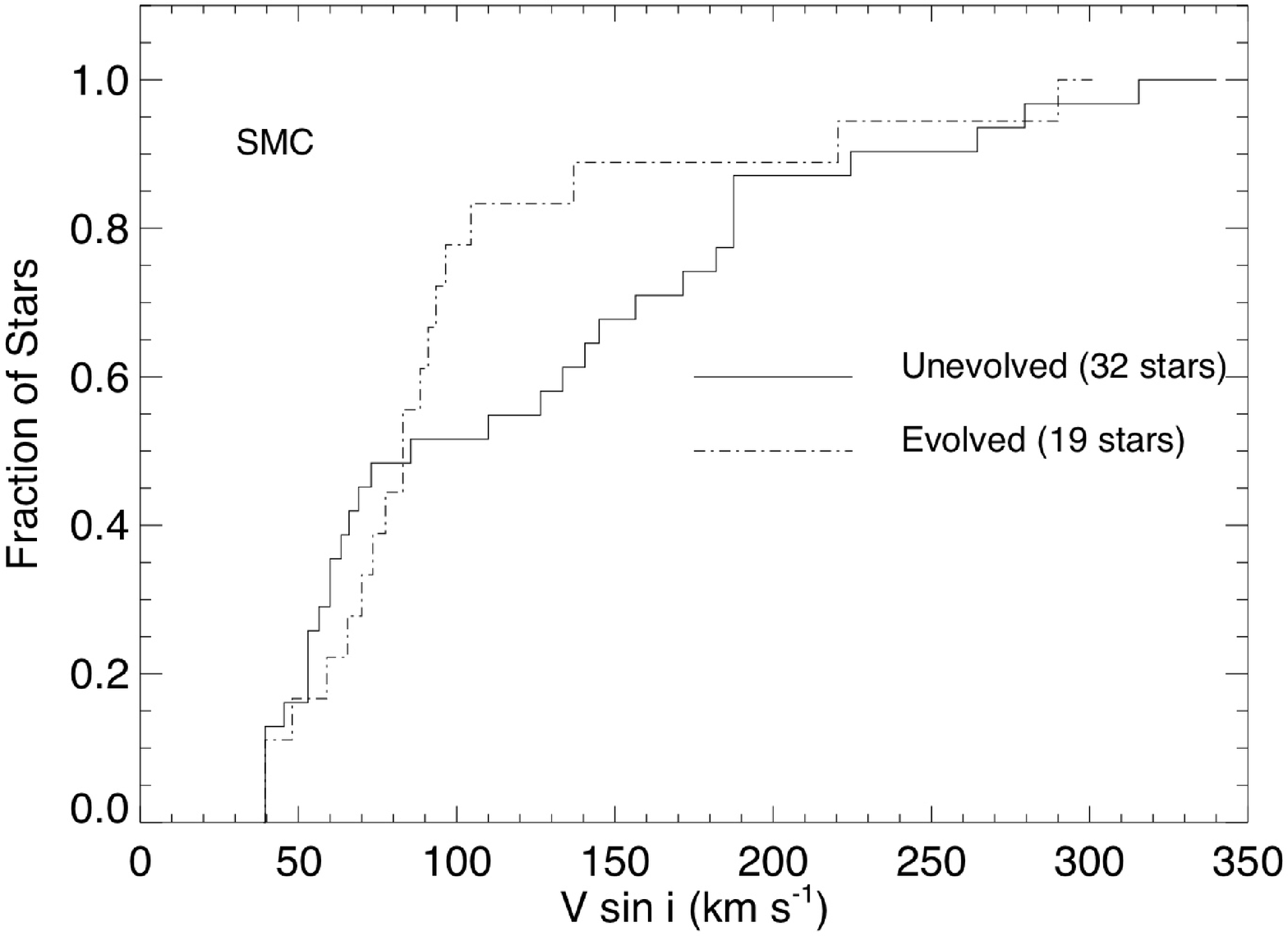}
\caption{Cumulative distribution functions of $V \sin i$ values for unevolved and evolved stars in the SMC.}
\label{f7}
\end{figure}

\placefigure{f8}      % Figure 8
\begin{figure}
\epsscale{1.0}
\plotone{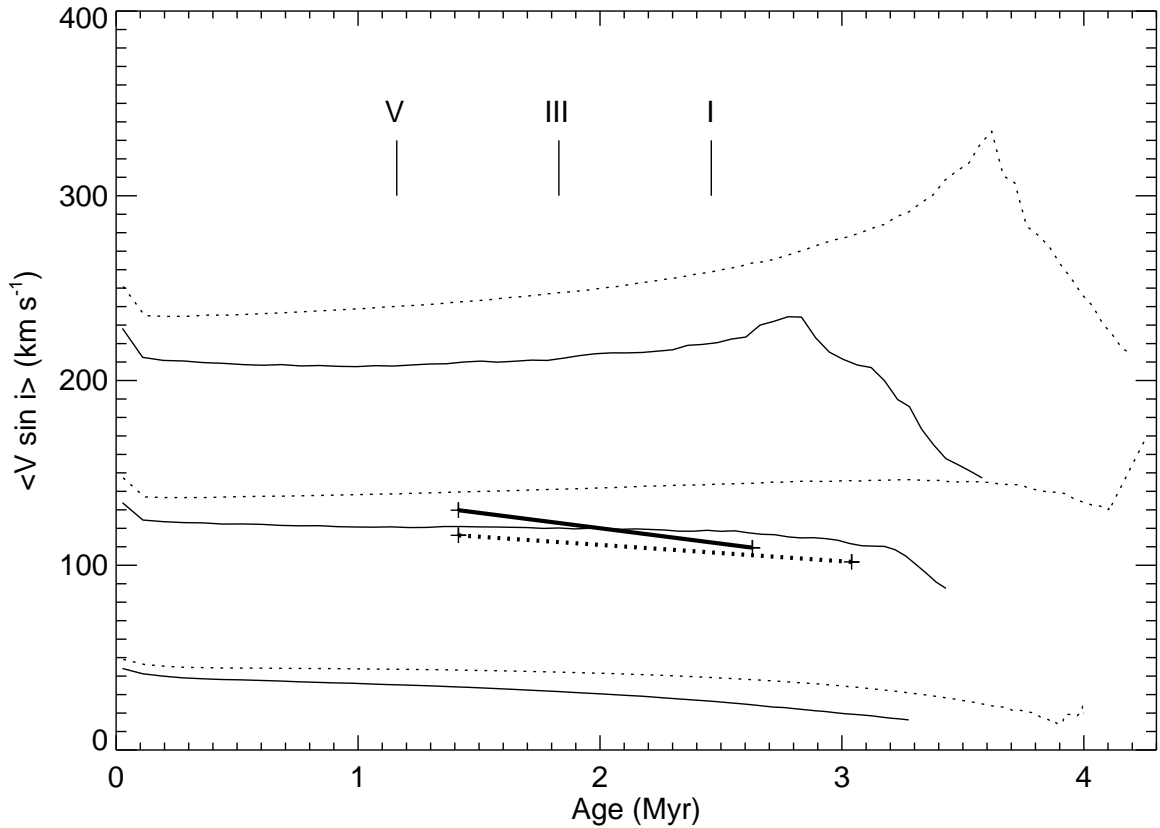}
\caption{Predicted changes in mean $V\sin i$ with age for
$60 M_\odot$ models from \citet{Eks08}.  The thin solid lines
show the variations for Galactic abundance stars with initial
ratios of equatorial to critical angular velocity of 0.5, 0.3, and
0.1 ({\it from top to bottom}), while the thin dotted lines show the
same for SMC abundance stars.  All these tracks are terminated at
an age where the star cools enough to become a B-supergiant.
The thick solid and dotted lines that connect plus signs show
the observed change in average $V\sin i$ between the unevolved
and evolved groups for the Galactic and SMC samples, respectively.}
\label{f8}
\end{figure}

\placefigure{f9}      % Figure 9
\begin{figure}
\epsscale{1.0}
\plotone{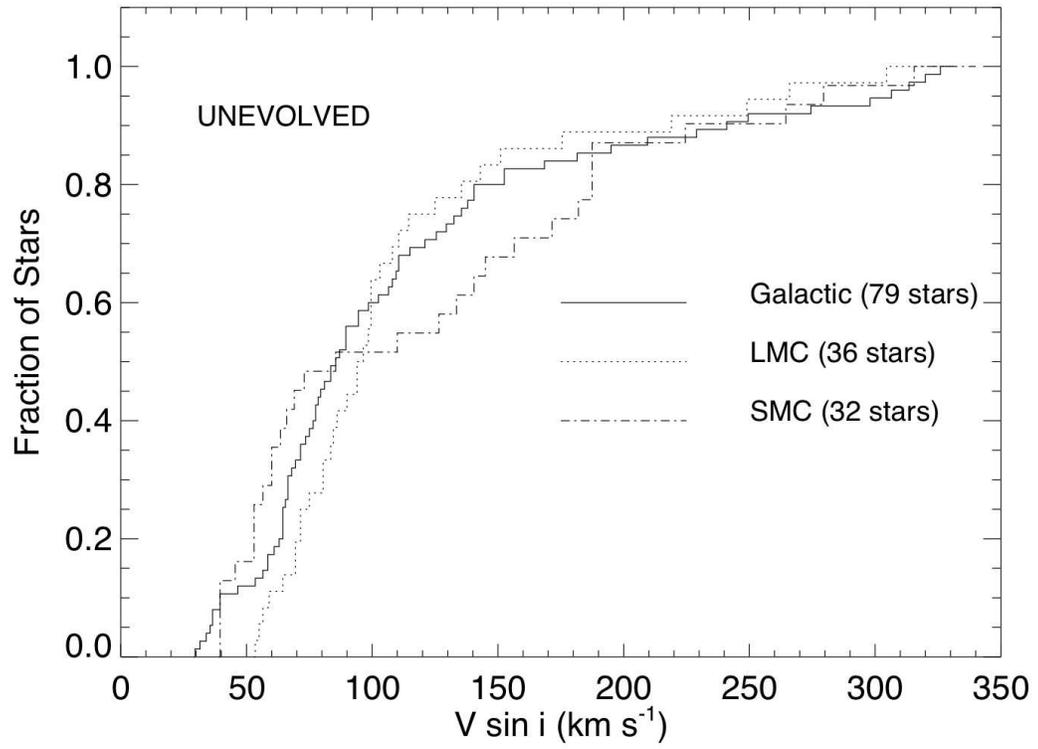}
\caption{Cumulative distribution functions of $V \sin i$ values for unevolved (luminosity classes IV - V) stars in the SMC, LMC and Galaxy.}
\label{f9}
\end{figure}

\placefigure{f10}      % Figure 10
\begin{figure}
\epsscale{1.0}
\plotone{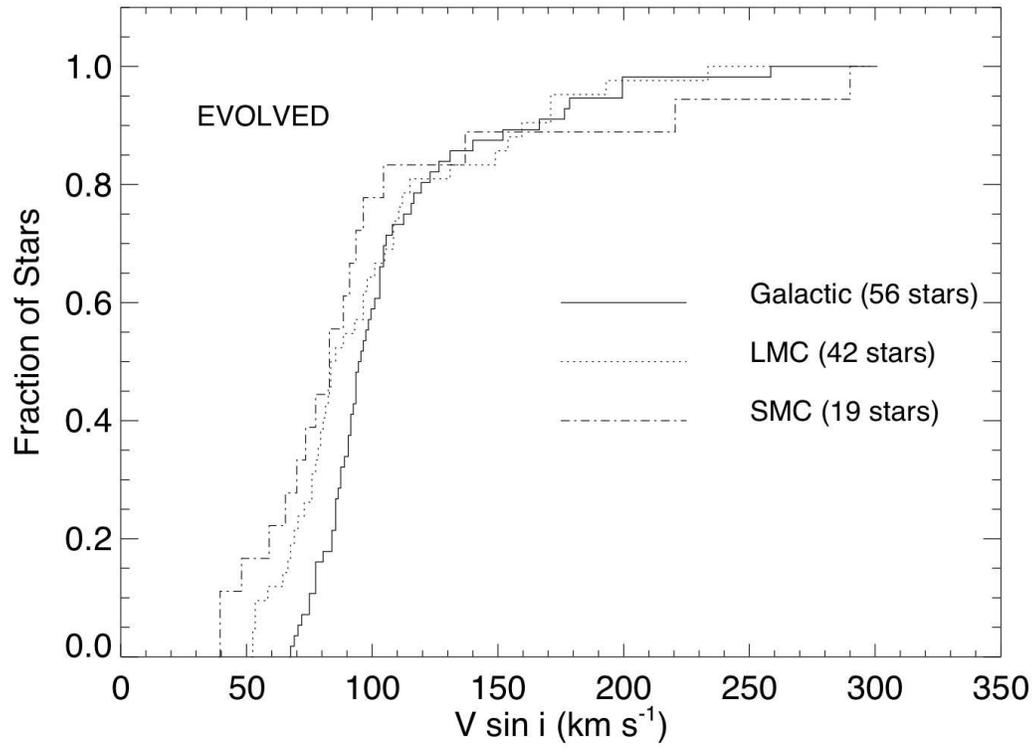}
\caption{Same as Figure 9 but for evolved (luminosity classes I -II) stars in the same three environments.}
\label{f10}
\end{figure}

\end{document}